\documentclass[prl,groupedaddress,superscriptaddress,twocolumn,amsmath,amssymb,a4paper]{revtex4}
\usepackage{geometry}
\usepackage[dvipsnames]{xcolor}
\usepackage{graphicx}
\usepackage{verbatim}
\usepackage{float}
\usepackage{graphics}
\usepackage{mathrsfs}
\usepackage{amssymb}
\usepackage{amsmath}
\usepackage{amsthm}
\usepackage{bm}
\usepackage{dsfont}
\usepackage{hyperref}
\usepackage{braket}
\usepackage{wrapfig}

\usepackage{amsbsy}

\begin{document}

\title{Machine Learning Non-Markovian Quantum Dynamics}

\author{I. A. Luchnikov}
\affiliation{Moscow Institute of Physics and Technology,
Institutskii Pereulok 9, Dolgoprudny, Moscow Region 141700,
Russia} \affiliation{Center for Energy Science and Technology,
Skolkovo Institute of Science and Technology, 3 Nobel Street,
Skolkovo, Moscow Region 121205, Russia}

\author{S. V. Vintskevich}
\affiliation{Moscow Institute of Physics and Technology,
Institutskii Pereulok 9, Dolgoprudny, Moscow Region 141700,
Russia}

\author{D. A. Grigoriev}
\affiliation{Moscow Institute of Physics and Technology,
Institutskii Pereulok 9, Dolgoprudny, Moscow Region 141700,
Russia}

\author{S. N. Filippov}
\affiliation{Moscow Institute of Physics and Technology,
Institutskii Pereulok 9, Dolgoprudny, Moscow Region 141700,
Russia} \affiliation{Valiev Institute of Physics and Technology of
Russian Academy of Sciences, Nakhimovskii Prospekt 34, Moscow
117218, Russia} \affiliation{Steklov Mathematical Institute of
Russian Academy of Sciences, Gubkina Street 8, Moscow 119991,
Russia}

\begin{abstract}
Machine learning methods have proved to be useful for the
recognition of patterns in statistical data. The measurement
outcomes are intrinsically random in quantum physics, however,
they do have a pattern when the measurements are performed
successively on an open quantum system. This pattern is due to the
system-environment interaction and contains information about the
relaxation rates as well as non-Markovian memory effects. Here we
develop a method to extract the information about the unknown
environment from a series of projective single-shot measurements
on the system (without resorting to the process tomography). The
method is based on embedding the non-Markovian system dynamics
into a Markovian dynamics of the system and the effective
reservoir of finite dimension. The generator of Markovian
embedding is learned by the maximum likelihood estimation. We
verify the method by comparing its prediction with an exactly
solvable non-Markovian dynamics. The developed algorithm to learn
unknown quantum environments enables one to efficiently control
and manipulate quantum systems.
\end{abstract}

\maketitle

\paragraph*{Introduction.---}
Quantum systems are never perfectly isolated which makes the study
of open quantum dynamics important for various disciplines
including solid-state physics~\cite{takahashi-2008}, quantum
chemistry~\cite{valkunas-2013}, quantum sensing~\cite{degen-2017},
quantum information transmission~\cite{wilde-2017}, and quantum
computing~\cite{nielsen-2000}. Open quantum dynamics is a result
of interaction between the system of interest and its environment.
It is usually assumed that the environment is an infinitely large
reservoir in statistical equilibrium, which has a well-defined
interaction with the system~\cite{schoeller-2018}. However, the
environments of many physical systems are rather complex and
structured~\cite{piilo-2011,cirac-2011,ma-2012,hoope-2012,yang-2013,hughes-2015,eisert-2015,cirac-2017,wittemer-2018,wang-2018,peng-2018,haase-2018,mascherpa-2019}.
A model of the system-environment interaction is often heuristic
and oversimplified (e.g., a harmonic environment), but even in
this case the analysis is rater complicated and requires some
elaborated analytical and numerical
methods~\cite{strathearn-2018,pollock-2019,altaisky-2017}. A
theoretical model may also neglect some additional sources of
decoherence and relaxation. The experimental analysis of the
environmental degrees of freedom is difficult because of their
inaccessibility in practice. In fact, one can only get some
information about the actual environment by probing the
system~\cite{paris-2018,bennink-2019}. Therefore, one faces an
important problem to learn the \textit{unknown} environment and
its interaction with the quantum system by probing and affecting
the system only.

This problem can be partly solved within the assumption of fast
bath relaxation, when the system density operator $\varrho_S$
experiences the semigroup dynamics $\varrho_S(t) = e^{{\cal L}_S
t} \varrho_S(0)$ with the Gorini-Kossakowski-Sudarshan-Lindblad
(GKSL) generator ${\cal L}_S$~\cite{gks-1976,lindblad-1976}. In
this case, the generator is reconstructed by performing a process
tomography of the channel $\Phi_S(t_1) = e^{{\cal L}_S t_1}$ for a
fixed time $t_1 > 0$~\cite{Tomography,howard-2006}. The actual
dynamics does not usually reduce to a semigroup
though~\cite{de-vega-2017,li-2018,fc-2018}. The problem of
learning the environment is mostly attributed to memory effects
accompanying the non-Markovian dynamics. In this case, one can
still resort to the process tomography of channels $\Phi_S(t_1)$,
$\Phi_S(t_2)$, $\ldots$, $\Phi_S(t_K)$ by preparing various
initial system states $\varrho_S(0)$ and performing different
measurements on the system at time moments $t_1 < t_2 < \ldots <
t_K$. This procedure is time consuming because one has to gather
enough statistics for all time moments (the total number of
required measurements is $K d_S^8 / \epsilon^2$ for a
$d_S$-dimensional quantum system and the accuracy $\epsilon$ of
statistical reconstruction~\cite{bogdanov-2013,haah-2017}).
Moreover, the tomographic reconstruction of each channel
$\Phi_S(t_{i})$ implies resetting the environment in the same
initial state after each measurement, which is difficult to
control in the experiment especially for a strong coupling between
the system and environment.

Recently proposed methods exploit the transfer tensor
techniques~\cite{Cerrilo-2014,TTM,TTMTomography} to learn the
Nakajima-Zwanzig equation~\cite{nakajima-1958,zwanzig-1960}
$\frac{d}{dt} \varrho_S(t) = \int_0^t {\cal K}(t-t')
\varrho_S(t')dt'$ and the recurrent neural networks~\cite{RNN} for
defining Lindblad operators and learning the convolutionless
master equation $\frac{d}{dt} \varrho_S(t) = {\cal L}_S(t)
\varrho_S(t)$. An implementation of the latter approach in
practice encounters the same difficulties related with the
necessity to perform state tomography at different time steps.

In this Letter, we develop a method to learn the effective
Markovian
embedding~\cite{MarkovEmb,xue-2015,xue-2017,campbell-2018,bennink-2019,luchnikov-2019}
for non-Markovian processes instead of learning the master
equation for the system ($S$). Within such an approach, the
environment is effectively divided into two parts: the first one
carries memory of the system and is responsible for non-Markovian
dynamics [\textit{effective reservoir} ($ER$)]; the second one is
memoryless and causes Markovian decoherence and dissipation of
$S+ER$. The system evolution reads
\begin{eqnarray}
\label{rho-system} \varrho_{S}(t) &=& {\rm tr}_{ER}
[\varrho_{S+ER}(t)], \\
\label{embedding} \frac{d\varrho_{S+ER}(t)}{dt} &=& {\cal
L}_{S+ER} [\varrho_{S+ER}(t)],
\end{eqnarray}

\noindent where the generator ${\cal L}_{S+ER}$ governs
dissipative and decoherence processes on the system and the
effective reservoir.

A division of the environment into two parts is similar to the
pseudomode method~\cite{imamoglu-1994,garraway-1997,mazzola-2009},
the reaction coordinate
model~\cite{iles-smith-2014,iles-smith-2016}, and the
non-Markovian core model~\cite{tamascelli-2018}, where one derives
a Markovian master equation in the GKSL form for the extended
system comprising the system and a finite number of auxiliary
modes. In spin-bosonic models, the Markovian embedding is
justified if the bath correlation function has exponentially
damped correlations~\cite{mascherpa-2019}. However, for power-law
bath correlation functions with long-range
tails~\cite{polyakov-2019} the number of auxiliary modes diverges,
which limits applicability of the Markovian embedding at a long
timescale.

\begin{figure}
\centering
\includegraphics[width=8.5cm]{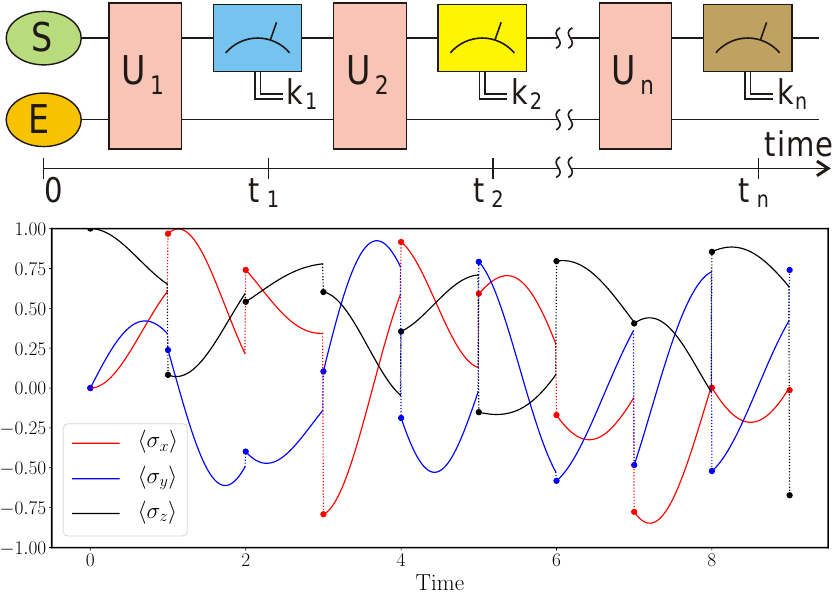}
\caption{(Top) Interventions into the open dynamics of the system
$S$ by projective measurements. Blocks $\{U_i\}$ depict the
interaction between $S$ and the actual environment ($E$) in
between the measurements. (Bottom) Example of the Bloch vector
evolution for a qubit system subjected to measurements in random
bases at time moments $t_i = i$. Circles correspond to the wave
function collapse.} \label{figure-1}
\end{figure}

The density operator $\varrho_S(t)$ is unaccessible in a single
measurement though, so any relevant information about the system
is only gained in a series of measurements. On the other hand,
measurement interventions into the system evolution complicate the
analysis due to the no-information-without-disturbance principle.
Consider a series of projective measurements performed on the
system at different times $t_1 < t_2 < \ldots < t_n$, with the
measurement basis being chosen randomly, see Fig.~\ref{figure-1}.
The measurement outcomes seem to be completely uninformative due
to the intrinsic probabilistic nature of quantum mechanics and the
wave function collapse at each measurement as an example in
Fig.~\ref{figure-1} suggests. However, such a series of
measurement outcomes does contain some information because the
outcomes at each time moment are not equiprobable but appear in
accordance with the Born rule. In this Letter, we demonstrate that
a sufficiently long series of measurement results has a
\emph{pattern} that can be recognized by a
machine~\cite{carleo-2019}. This is a sharp distinction from
conventional tomographic approaches based on numerous repetitions
of identical experiments to gather enough statistics.

Our algorithm maximizes the likelihood of observed measurement
outcomes and provides the generator ${\cal L}_{S+ER}$ for any
fixed dimension of the effective reservoir $d_{ER}$, which is a
hyperparameter. Computationally, the optimal $d_{ER}$ corresponds
to the maximal likelihood on the validation set, which prevents
overfitting~\cite{supplemental}. Physically, the sufficient value
of $d_{ER}$ can be estimated through a reduced set of parameters:
the system-environment coupling strength, the reservoir
correlation time, the cutoff frequency of the spectral function,
and the system's number of degrees of freedom interacting with the
environment~\cite{luchnikov-2019,supplemental}. Alternatively,
$d_{ER}$ can also be estimated via the ensemble learning
method~\cite{shrapnel-2018}.

\begin{figure}
\centering
\includegraphics[width=8.5cm]{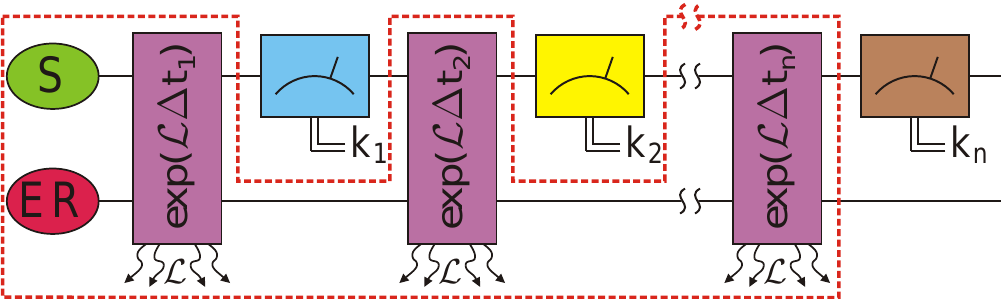}
\caption{Markovian embedding of the open dynamics: $S$ and
effective reservoir $ER$ experience semigroup dynamics with the
generator ${\cal L}$; $\Delta t_i = t_i - t_{i-1}$. The process
tensor is depicted by the dotted line.} \label{figure-2}
\end{figure}

If the system evolution is Markovian ($d_{ER}=1$), then the result
of measurement at time $t_k$ depends on the measurement outcome at
time $t_{k-1}$ only and does not depend on results of earlier
measurements at times $t_{k-2}, t_{k-3},
\ldots$~\cite{lindblad-1979}. Instead, the non-Markovian dynamics
is accompanied by correlations in the measurement
outcomes~\cite{lindblad-1979,pollock-2018,modi-2018,budini-2018,taranto-2019},
which can be analyzed via the process matrix~\cite{costa-2016} and
the process tensor~\cite{milz-2017}. The process tensor is a
particular form of a quantum network~\cite{chiribella-2009}, which
is defined through the generator ${\cal L}_{S+ER}$ in our model,
see Fig.~\ref{figure-2}. The reconstruction of a general process
tensor requires exponentially many projective
measurements~\cite{milz-2018}. However, the process tensor has a
peculiar form in our model and depends on the generator ${\cal
L}_{S+ER}$ only, so it can be reconstructed by maximizing the
likelihood of getting the observed outcomes for a single series of
measurements without resorting to the full quantum tomography.

\paragraph*{Likelihood function and its gradient.---}
Suppose the experimental setup allows for projective measurements
of the system at times $t_i = i \tau$, $i=1,\ldots,n$, with the
measurement basis $\{|\varphi_k^{(i)}\rangle\}_{k=1}^{d_S}$ being
randomly chosen at each time moment $t_i$. Observation of the
particular measurement outcome $k_i$ transforms the system state
into $|\varphi_{k_i}^{(i)}\rangle \langle\varphi_{k_i}^{(i)}|$.
Denote $E_i = |\varphi_{k_i}^{(i)}\rangle
\langle\varphi_{k_i}^{(i)}| \otimes I_{ER}$ the projector acting
on the system and effective reservoir. The collection of
projectors $\{E_i\}_{i=1}^n$ is the training \textit{dataset} that
feeds the learning algorithm.

A superoperator $\Phi = \exp\left(\tau {\cal L}_{S+ER}\right)$
governs the system and the effective reservoir evolution in
between two sequential measurements. The probability to get the
particular sequence of measurement outcomes $\{k_i\}_{i=1}^n$ (the
data $\{E_i\}_{i=1}^n$) equals~\cite{luchnikov-2017}
\begin{equation} \label{likelihood-Phi}
p = {\rm tr} \left\{ E_n \ldots \Phi\big[ E_1
\Phi[\varrho_{S+ER}(0)] E_1 \big] \ldots E_n \right\}.
\end{equation}

\begin{figure}
\centering
\includegraphics[width=8.5cm]{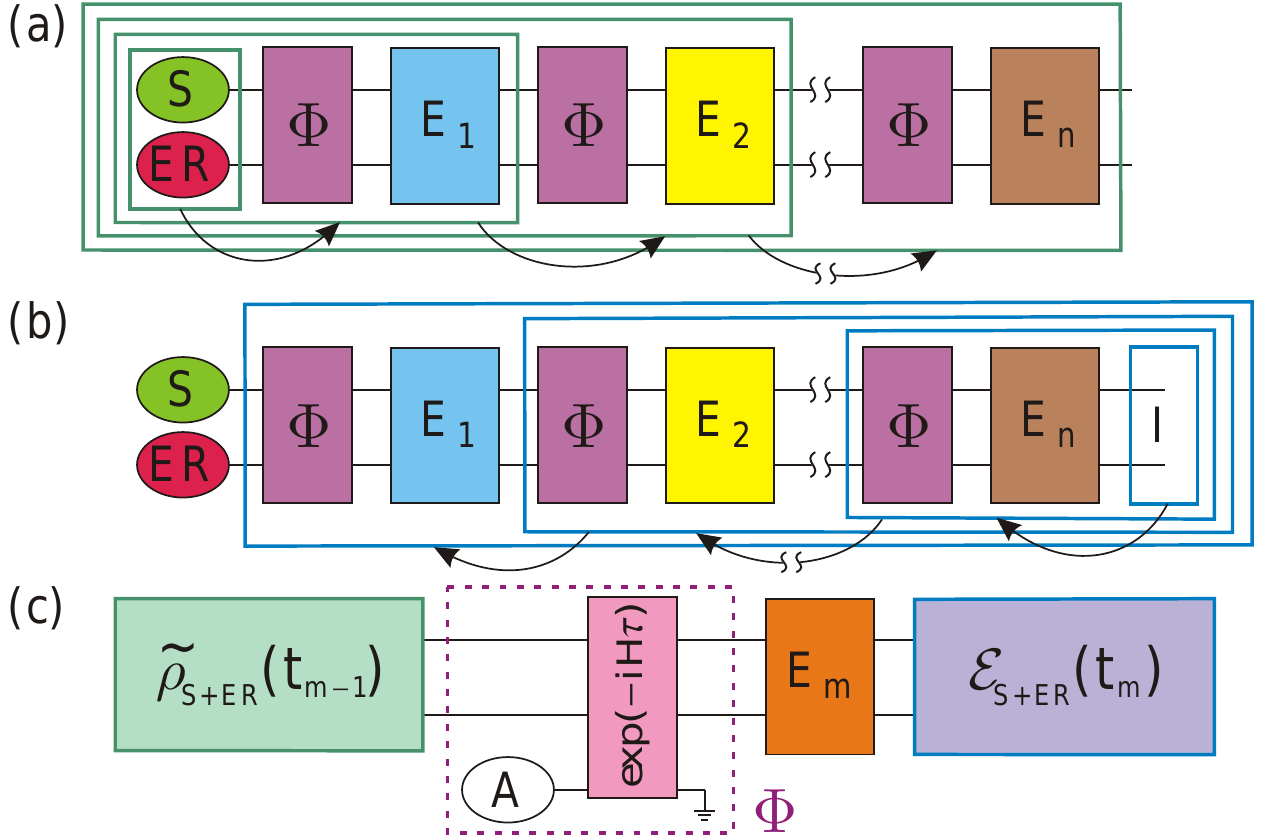}
\caption{(a) Forward propagation for subnormalized density
operators $\widetilde{\varrho}_{S+ER}(t_i)$. (b) Backward
propagation for effects ${\cal E}_{S+ER}(t_i)$. (c) Likelihood
functional in Eq.~\eqref{likelihood} and the Stinespring
dilation~\eqref{Phi} for $\Phi$ (dotted line).} \label{figure-3}
\end{figure}

The likelihood~\eqref{likelihood-Phi} admits alternative useful
forms. Let $\Phi^{\dag}$ be dual to $\Phi$~\cite{comment}, then
one can split Eq.~\eqref{likelihood-Phi} after $m$th measurement
and get $p = {\rm tr} \big[ \widetilde{\varrho}_{S+ER}(t_m) {\cal
E}_{S+ER}(t_m) \big]$, where the recurrence relation
$\widetilde{\varrho}_{S+ER}(t_{i+1}) = E_i
\Phi[\widetilde{\varrho}_{S+ER}(t_i)] E_i$ with
$\widetilde{\varrho}_{S+ER}(0) = \varrho_{S+ER}(0)$ defines the
\textit{forward} propagation of the subnormalized density operator
along the tensor network in Fig.~\ref{figure-3}(a) and the
recurrence relation ${\cal E}_{S+ER}(t_{i-1}) = \Phi^{\dag}[E_i
{\cal E}_{S+ER}(t_i) E_i]$ with ${\cal E}_{S+ER}(t_n) = I_{S+ER}$
defines the \textit{backward} propagation~\cite{molmer-2013} of
effects in the Heisenberg picture along the tensor network in
Fig.~\ref{figure-3}(b). This leads to a ``sandwich'' formula
\begin{equation} \label{likelihood}
p =  {\rm tr} \big\{ \Phi[ \widetilde{\varrho}_{S+ER}(t_{m-1}) ]
E_{m} {\cal E}_{S+ER}(t_{m}) E_{m} \big\},
\end{equation}

\noindent which is valid for all $m=1,\ldots,n$, see
Fig.~\ref{figure-3}(c).

The likelihood function is to be maximized over parameters of the
generator ${\cal L}_{S+ER}$ defining $\Phi = \exp(\tau {\cal
L}_{S+ER})$. Such a maximization is the most common approach in
supervised machine learning~\cite{Bishop}. The problem is that not
every generator ${\cal L}_{S+ER}$ defines a legitimate (completely
positive and trace preserving) map $\Phi$. To overcome this
obstacle and simplify the implementation of the gradient ascent
method~\cite{CVX}, we use the Stinespring dilation for the channel
$\Phi$ [see, e.g.,~\cite{Holevo} and Fig.~\ref{figure-3}(c)]:
\begin{equation} \label{Phi}
\Phi[\varrho_{S+ER}] = {\rm tr}_{A} \left[ U(H) \, \varrho_{S+ER}
\otimes \varrho_{A} \, U^{\dag}(H) \right],
\end{equation}

\noindent where $\varrho_A$ is a \textit{fixed} pure state of the
$d_A$-dimensional ancilla ($A$), $d_A = (d_S d_{ER})^2$, $U(H) =
\exp(-i H \tau)$ is a unitary evolution operator, and $H$ is the
effective Hamiltonian of $S+ER+A$. Eq.~\eqref{Phi} guarantees
$\Phi$ is completely positive and trace preserving provided $H$ is
Hermitian. The ancillary operator $\varrho_A$ plays the role of a
renewable subenvironment in quantum collision
models~\cite{rau-1963,scarani-2002,filippov-2017} and memoryless
(Markovian) part of the environment~\cite{shrapnel-2018}.

Because of the Stinespring dilation, the likelihood function is
now to be maximized over parameters of the effective Hamiltonian,
i.e., matrix elements $H_{\mu\nu} = \langle \mu | H | \nu \rangle$
of $H$ in some computational basis $\{ | \mu \rangle
\}_{\mu=1}^{d_S d_{ER} d_A}$. This means that parameters
$H_{\mu\nu}$ are iteratively adjusted in the direction of the
gradient of the logarithmic likelihood $g_{\mu\nu} =
\frac{\partial \log p}{\partial H_{\mu\nu}}$. Since the likelihood
function is the $n$-degree monomial with respect to both operators
$U(H)$ and $U^{\dag}(H)$, we readily get~\cite{supplemental}
\begin{eqnarray} \label{log-gradient}
&& \!\!\!\!\!\!\!\!\!\! g_{\mu\nu} =
\frac{1}{p} \sum_{m=1}^n {\rm tr} \bigg\{ [E_{m} \, {\cal E}_{S+ER}(t_{m}) \, E_{m}] \otimes I_A \nonumber\\
&& \!\!\!\!\!\!\!\!\!\! \times \bigg[ \frac{\partial
U(H)}{\partial H_{\mu\nu}} \,\,
\widetilde{\varrho}_{S+ER}(t_{m-1}) \otimes \varrho_A \,\,
U^{\dag}(H) + {\rm H.c.} \bigg] \bigg\}, \quad
\end{eqnarray}

\noindent where the derivative $\frac{\partial U(H)}{\partial
H_{\mu\nu}}$ is expressed through the spectral decomposition
$H=\sum_k \lambda_k |\psi_k\rangle \langle\psi_k|$
as~\cite{supplemental}
\begin{equation} \label{U-derivative}
\frac{\partial U(H)}{\partial H_{\mu\nu}} = \sum_{k,l}
\frac{e^{-i\lambda_k\tau} - e^{-i\lambda_l \tau}}{\lambda_k -
\lambda_l} \langle \psi_k | \mu \rangle \langle \nu | \psi_l
\rangle \, |\psi_k\rangle \langle\psi_l|.
\end{equation}

\noindent Keeping in a computer memory the operators
$\widetilde{\varrho}_{S+ER}(t_i)$ and ${\cal E}_{S+ER}(t_i)$ for
forward and backward propagations, respectively, we efficiently
calculate the gradient in $O(n)$ steps. Since $\log p$ is a highly
nonlinear and nonconvex function with respect to parameters
$H_{\mu\nu}$, its optimization is accompanied with overcoming the
convergence to local extremums and the slow convergence rate. In
what follows, we use techniques that were shown to perform well in
such nonconvex optimization problems as neural network
learning~\cite{jain-2017}.

\paragraph*{Learning algorithm.---}
The learning algorithm, which estimates the generator ${\cal
L}_{S+ER}$ based on the training dataset $\{E_i\}_{i=1}^n$, is as
follows~\cite{grigoriev_github}:
\begin{enumerate}

\item Fix the hyperparameter $d_{ER}$. Initialize the model by
randomly choosing the factorized state
$\varrho_{S+ER}(0)=\varrho_S(0) \otimes \varrho_{ER}(0)$ and the
factorized Hamiltonian $H=H_{S+ER} \otimes I_{A}$.

\item \label{operators-in-memory} Calculate the
forward-propagation operators
$\{\widetilde{\varrho}_{S+ER}(t_i)\}_{i=1}^n$ and the
backward-propagation operators $\{{\cal
E}_{S+ER}(t_i)\}_{i=0}^{n-1}$.

\item \label{likelihood-in-algorithm} Calculate the
likelihood~\eqref{likelihood}.

\item \label{U-derivative-in-algorithm} Find the spectral
decomposition of the $d_S d_{ER} d_A$-dimensional operator $H$ and
calculate $\frac{\partial U(H)}{\partial H_{\mu\nu}}$ via
\eqref{U-derivative}.

\item Estimate the gradient~\eqref{log-gradient} via a batch of
summands and results of items \ref{operators-in-memory},
\ref{likelihood-in-algorithm}, \ref{U-derivative-in-algorithm}.

\item \label{Hamiltonian-update} Feed the estimated gradient to an
advanced optimization method [e.g., the adaptive moment estimation
(Adam) algorithm~\cite{Adam}] and get the increment $\Delta H$.

\item Update the Hamiltonian $H \rightarrow H + \Delta H$ and
repeat items \ref{operators-in-memory}--\ref{Hamiltonian-update}
until the likelihood converges.

\item Make use of the final update of $H$ to find the channel
$\Phi$ and the generator ${\cal L}_{S+ER} = \frac{1}{\tau} \ln
\Phi$.

\end{enumerate}

\paragraph*{Synthetic data generation.---}
We apply the learning algorithm above to the \textit{in silico}
training set $\{E_i\}_{i=1}^n$ generated in a non-Markovian
composite bipartite collision model~\cite{lorenzo-2017}. We
consider a bipartite system $S+S_1$ composed of the very open
qubit system under study $S$ and one auxiliary qubit system $S_1$.
The bipartite system successively interacts with identical
subenvironments during some collision time~\cite{supplemental}.
Such a model is quite rich and describes, e.g., a qubit subject to
random telegraph noise. The benefit of this model is that the
measurement interventions into the system evolution are explicitly
taken into account~\cite{supplemental}.

\begin{figure}
\centering
\includegraphics[width=8.5cm]{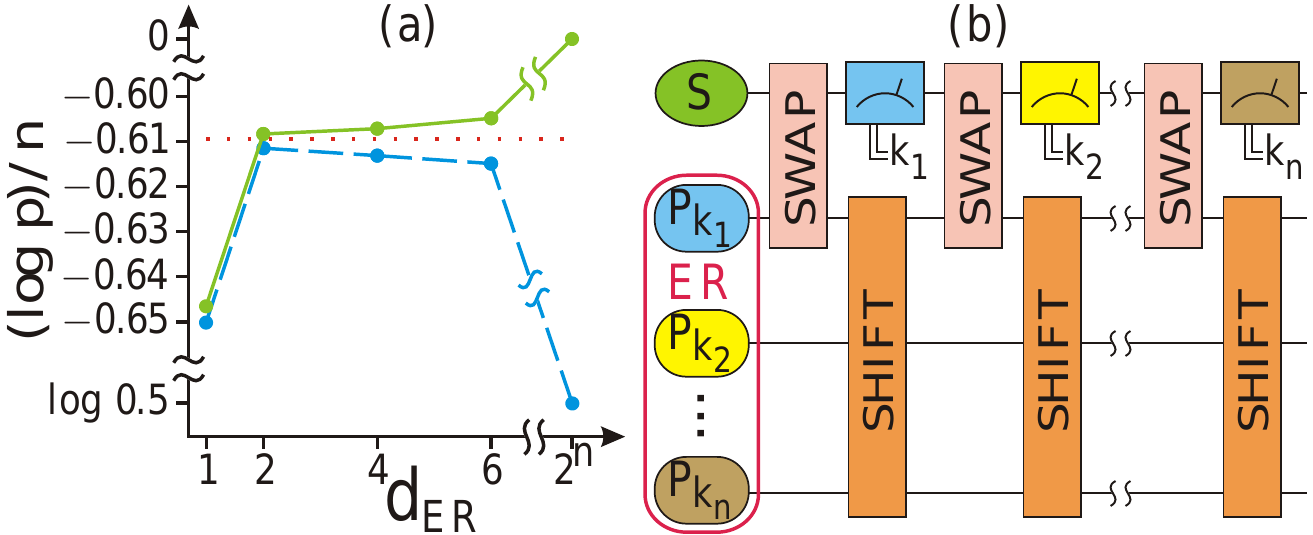}
\caption{(a) Logarithmic likelihood per measurement vs dimension
of the effective reservoir for the training set $\{E_i\}_{i=1}^n$
(solid line) and the validation set $\{E_i\}_{i=n+1}^{2n}$ (dashed
line). Theoretical prediction for the generated data is depicted
by a dotted line. (b) Ultimate overfitting with the exponentially
big effective reservoir composed of the projectors observed, swap
gates, and the shift operator $i  \rightarrow i-1, 1 \rightarrow
n$ for subenvironments.} \label{figure-4}
\end{figure}

\paragraph*{Validation.---}
We run the learning algorithm for various values of the
hyperparameter $d_{ER} = 1,2,4,6$ on the generated training set
$\{E_i\}_{i=1}^n$, $n=10^5$~\cite{supplemental}. The value $d_{ER}
= 1$ corresponds to the best Markovian approximation for the
dynamics that is most compatible with the observed measurement
outcomes. However, the likelihood for $d_{ER} = 1$ is less than
that for non-Markovian models with $d_{ER} \geq 2$, see
Fig.~\ref{figure-4}(a). The greater $d_{ER}$, the wider the
complexity class of possible dynamics~\cite{luchnikov-2019}. If
$d_{ER} = d_S^n$, then any series of projectors $\{E_i\}_{i=1}^n$
can be perfectly reconstructed with the likelihood
$p(\{E_i\}_{i=1}^n) = 1$, which is an ultimate case of
overfitting~\cite{supplemental}, see Fig.~\ref{figure-4}(b). The
maximally achieved values of the logarithmic likelihood $\log
p(\{E_i\}_{i=1}^n)$ on the training set monotonically increase
with the increase of $d_{ER}$. To avoid overfitting, we calculate
the likelihood~\eqref{likelihood-Phi} on a separate
\textit{validation set} of projectors $\{E_i\}_{i=n+1}^{2n}$.
Fig.~\ref{figure-4}(a) shows that, for the data analyzed, the
logarithmic likelihood $\log p(\{E_i\}_{i=n+1}^{2n})$ on the
validation set increases up to $d_{ER} = 2$ and then diminishes.
The Markovian embedding with $d_{ER} = 2$ is the simplest model
that is the most compatible with the observed series of
measurement outcomes. This is an expected result because we used
the synthetic data generated within a collision model with qubits,
$d_{S_1} = 2$. For real experimental data, the hyperparameter
$d_{ER}$ is tuned in such a way that the likelihood on the
validation set achieves its maximum. Tuning is reasonable to
perform in the vicinity of the physical estimate for $d_{ER}$
derived in Ref.~\cite{luchnikov-2019}.

\begin{figure}
\centering
\includegraphics[width=8.5cm]{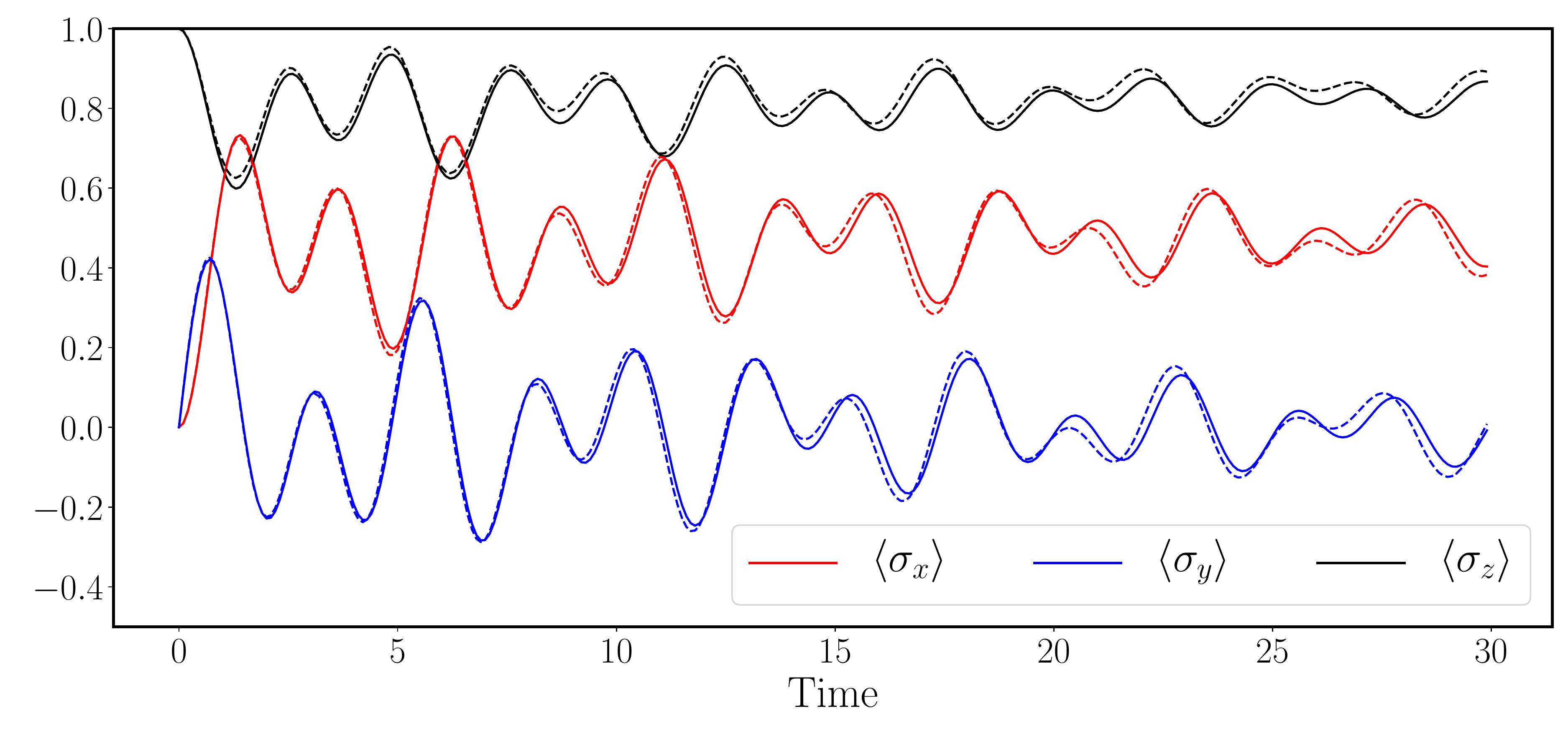}
\caption{Bloch vector components $\langle \sigma_i(t) \rangle =
{\rm tr}[\varrho_S(t) \sigma_i]$ vs dimensionless time for the
exact dynamics (solid line) and the learning-based prediction
(dotted line).} \label{figure-5}
\end{figure}

\paragraph*{Results.---} With the estimated
generator ${\cal L}_{S+ER}$ at hand, we predict the open system
dynamics $\varrho_S(t)$ by Eqs.~\eqref{rho-system} and
\eqref{embedding} and compare it with the exact theoretical model
(with no measurement interventions). The missing initial state of
the effective reservoir is chosen to be the equilibrium state
${\rm tr}_S[\varrho_{S+ER}^{\infty}]$ such that ${\cal
L}_{S+ER}[\varrho_{S+ER}^{\infty}] = 0$. The results are depicted
in Fig.~\ref{figure-5}. Good agreement between the estimated
dynamics and the exact one demonstrates that the presented
learning algorithm actually extracts useful information from the
correlation pattern in a sequence of measurements on the open
quantum system.

The quality of the estimated dynamics is assessed in two ways. (i)
If the exact dynamics $\Phi_S(t)$ is known, we calculate the
distinguishability between the estimated dynamics and the exact
one, then average over time moments within the interval $[0,T]$.
The result is $\varepsilon = 0.03$ for $T =
50$~\cite{supplemental}. (ii) If the exact solution is not known,
the quality of the estimated dynamics is assessed within the
variational Bayesian inference approach. This approach yields
$\varepsilon = 0.05$ for $T = 50$ and the standard deviation 0.025
for matrix elements of the estimated density operator
$\varrho_S(T)$~\cite{supplemental}.

The average error in estimating the discretized process
$\{\Phi_S(t_i)\}_{i=1}^K$ scales as $1/\sqrt{n}$ and is
essentially independent of $K$ in the proposed
algorithm~\cite{supplemental} because all the channels
$\{\exp({\cal L}_{E+ER}\Delta t_i)\}_{i=1}^K$ in the process
tensor in Fig.~\ref{figure-1} are defined by a single generator
${\cal L}_{S+ER}$ independent of time moments $\{t_i\}_{i=1}^K$
(parameter sharing). On the other hand, the full process
tomography yields the error scaling as $\sqrt{K/n}$ with the same
total number of measurements $n$~\cite{supplemental}. Therefore,
the proposed method is $\sqrt{K}$ times more efficient as compared
to the full process tomography for large $K$.

\begin{figure}
\centering
\includegraphics[width=8.5cm]{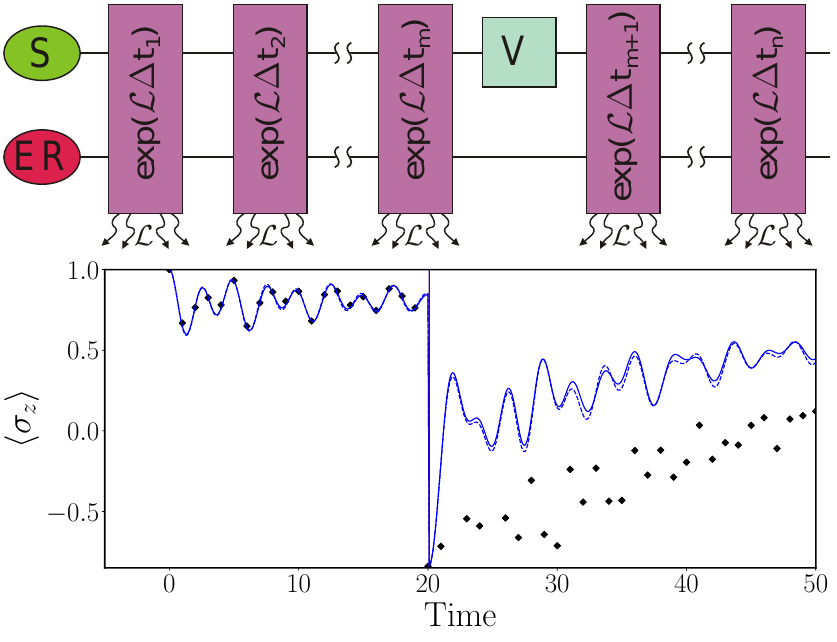}
\caption{Compatibility of the process tensor formalism with a
coherent control gate $V$ applied to the system (top). Example of
non-Markovian qubit dynamics with a quick control gate $V =
\sigma_x$ applied at $t'=20$ (bottom): exact solution (solid
line), estimated solution within the Markovian embedding approach
(dotted line), solution within the full process tomography
approach (dots).} \label{figure-6}
\end{figure}

Importantly, the formalism of Markovian embedding is compatible
with a control operation on system $S$, say, a quick unitary
transformation $\varrho_S(t') \rightarrow V\varrho_S(t') V^{\dag}$
at time moment $t'$. After the operation, $\varrho_S(t) = {\rm
tr}_{ER} \big\{ \exp[(t-t'){\cal L}_{S+ER}]
\varrho_{S+ER}(t')\big\}$. The result is in good agreement with
the exact dynamics (Fig.~\ref{figure-6}), thus opening an avenue
toward efficient control and manipulation of non-Markovian quantum
systems. In contrast, the conventional process tomography cannot
take such a control operation into account: its prediction
$\Phi_S(t)\Phi_S(t')^{-1}[V\varrho_S(t') V^{\dag}]$ differs from
$\varrho_S(t)$ because of the system-environment
correlations~\cite{supplemental,gessner-2011,rivas-2014,milz-2019},
see Fig.~\ref{figure-6}.

\paragraph*{Conclusions.---}
We proposed a method to learn the Markovian embedding for
non-Markovian quantum evolution. The primary information needed is
the outcomes of successive projective measurements on the system.
Correlations in the measurements at different times indicate
non-Markovianity and allow for the reconstruction of memory
effects. The decay of correlations between spaced-in-time
measurements enables the reconstruction of relaxation effects.
Both memory and relaxation phenomena are taken into account by the
generator ${\cal L}_{S+ER}$ acting on the system and the effective
reservoir of finite dimension. Our algorithm estimates ${\cal
L}_{S+ER}$ and does not exploit the full tomography of either
states or processes. Learnability of the algorithm is tested on a
dataset for the non-Markovian qubit dynamics. The presented
approach enables to take control on the system into consideration,
which is impossible with conventional tomographic techniques.

\begin{acknowledgements}
The authors thank Peter Sta\v{n}o for useful comments.
Conceptualization, implementation, and validation of the learning
algorithm is supported by the Russian Foundation for Basic
Research under Project Nos. 18-37-00282 and 18-37-20073 and is
performed in Moscow Institute of Physics and Technology, Skolkovo
Institute of Science and Technology, and Valiev Institute of
Physics and Technology of Russian Academy of Sciences, where
S.N.F. was partially supported by Program No. 0066-2019-0005 of
the Russian Ministry of Science and Higher Education. Synthetic
data generation is performed in Moscow Institute of Physics and
Technology, where I.A.L. and S.N.F. were partially supported by
the Foundation for the Advancement of Theoretical Physics and
Mathematics ``BASIS'' under Grant No. 19-1-2-66-1. The study of
the likelihood function and quantum control is supported by the
Russian Science Foundation under Project No. 17-11-01388 and is
performed in Steklov Mathematical Institute of Russian Academy of
Sciences.
\end{acknowledgements}


\newpage

\begin{widetext}

\begin{center}
{\large \bf SUPPLEMENTAL MATERIAL}
\end{center}

\subsection{Derivation of the likelihood function gradient}

The likelihood function $p({E_i}_{i=1}^n)$ can be rewritten in
many alternative ways with the help of the forward propagation
operators $\widetilde{\varrho}_{S+ER}(t_i)$ and the backward
propagation operators ${\cal E}_{S+ER}(t_i)$. In fact,
$\widetilde{\varrho}_{S+ER}(t_i)$ is the subnormalized state of
$S+ER$ at time moment $t_i$ such that ${\rm
tr}[\widetilde{\varrho}_{S+ER}(t_m)] = p(\{E_i\}_{i=1}^m)$. The
total likelihood for $n$ outcomes equals $p(\{E_i\}_{i=1}^n) =
{\rm tr}[\widetilde{\varrho}_{S+ER}(t_n)] = {\rm
tr}[\widetilde{\varrho}_{S+ER}(t_n) I_{S+ER}] = {\rm
tr}[\widetilde{\varrho}_{S+ER}(t_n) {\cal E}_{S+ER}(t_n)]$, where
we have introduced the ``initial'' condition for backward
propagation ${\cal E}_{S+ER}(t_n) = I_{S+ER}$.

Forward propagation is given by the recurrence relation
\begin{equation} \label{recurrent-forward-supplemental}
\widetilde{\varrho}_{S+ER}(t_{i+1}) = E_i
\Phi[\widetilde{\varrho}_{S+ER}(t_i)] E_i = {\rm tr}_A \big[ E_i
\otimes I_A \, U(H) \, \widetilde{\varrho}_{S+ER}(t_i) \otimes
\varrho_A U^{\dag}(H)\, E_i \otimes I_A \big]
\end{equation}

\noindent with $\widetilde{\varrho}_{S+ER}(0) =
\varrho_{S+ER}(0)$. The last equality
in~\eqref{recurrent-forward-supplemental} is due to the
Stinespring dilation for the channel $\Phi$ exploiting an ancilla
$A$ of dimension $d_A = (d_S d_{ER})^2$ (the maximum Kraus rank
for the channel $\Phi$~\cite{Holevo}). Tensor diagram for
Eq.~\eqref{recurrent-forward-supplemental} is depicted in
Fig.~\ref{figure-1-supplemental}. Ancillary systems $A$ play the
role of particles colliding with $S+ER$ (see,
e.g.,~\cite{fsp-2020}).

\begin{figure}[b]
\centering
\includegraphics[height=4cm]{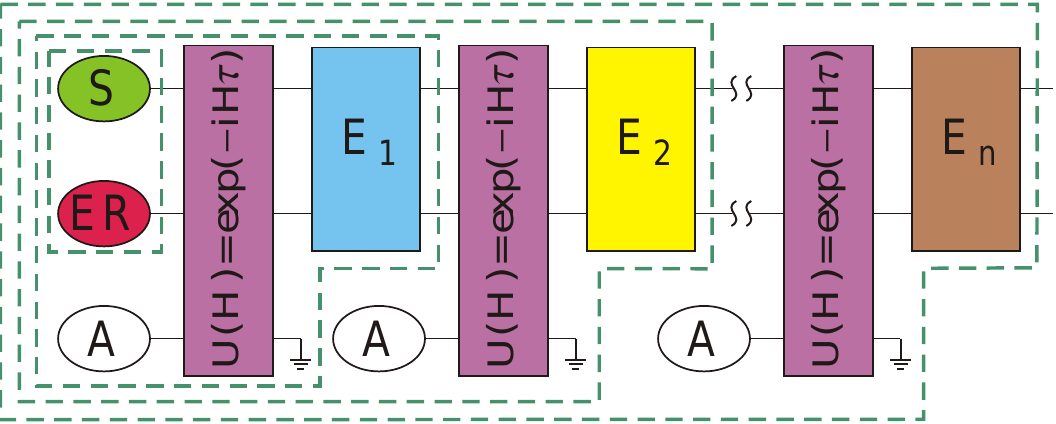}
\caption{Extended tensor network for the forward propagation.}
\label{figure-1-supplemental}
\end{figure}

Similarly, the operator ${\cal E}_{S+ER}(t_m)$ propagates
backward, i.e., in the Heisenberg picture, and is given by the
recurrence relation
\begin{equation} \label{recurrent-backward-supplemental}
{\cal E}_{S+ER}(t_{i-1}) = \Phi^{\dag}[E_i {\cal E}_{S+ER}(t_i)
E_i] = {\rm tr}_A \big[ U^{\dag}(H) \, ( E_{i} {\cal
E}_{S+ER}(t_{i}) E_{i} \otimes I_A ) U(H) \, (I_{S+ER} \otimes
\varrho_A) \big]
\end{equation}

\noindent with ${\cal E}_{S+ER}(t_n) = I_{S+ER}$. The last
equality in~\eqref{recurrent-backward-supplemental} explicitly
uses the Stinesping dilation for the dual channel $\Phi^{\dag}$.
Tensor diagram for Eq.~\eqref{recurrent-backward-supplemental} is
depicted in Fig.~\ref{figure-2-supplemental}.

\begin{figure}[t]
\centering
\includegraphics[height=4cm]{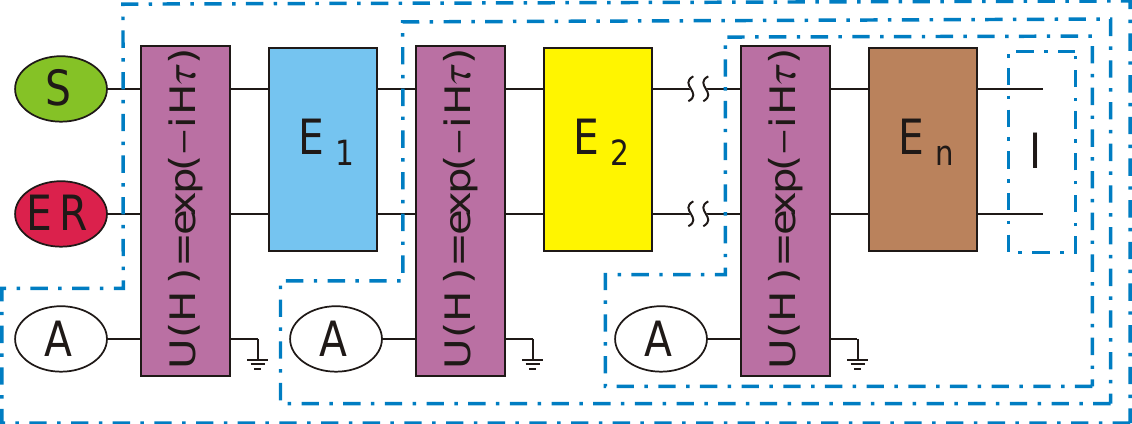}
\caption{Extended tensor network for the backward propagation.}
\label{figure-2-supplemental}
\end{figure}

Merging the forward and backward propagations at a fixed time
$t_m$, we express the likelihood function in many various though
equivalent forms, namely,
\begin{eqnarray} \label{likelihood-supplemental}
p(\{E_i\}_{i=1}^n | H) & = & {\rm tr}\left[ \widetilde{\varrho}_{S+ER}(t_m) {\cal E}_{S+ER}(t_m) \right] \quad \forall m=0,\ldots,n, \ t_0=0 \nonumber\\
& = & {\rm tr}\big[ U(H) \, \widetilde{\varrho}_{S+ER}(t_{m-1})
\otimes \varrho_A \, U^{\dag}(H) \,\, E_{m} \, {\cal
E}_{S+ER}(t_{m}) \, E_{m} \otimes I_A \big] \quad \forall
m=1,\ldots,n.
\end{eqnarray}

\noindent The latter expression~\eqref{likelihood-supplemental} is
a ``sandwich'' composed of the forward propogation till time
$t_{m-1}$ [the state $\widetilde{\varrho}_{S+ER}(t_{m-1})$], the
backward propagation till time $t_{m}$ [the operator ${\cal
E}_{S+ER}(t_{m})$], and the unitary transformation $U(H) \, \cdot
\, U^{\dag}(H)$ followed by the $m$th measurement in between.
Since the likelihood function is the $n$-degree monomial with
respect to both operators $U(H)$ and $U^{\dag}(H)$, we readily
express its gradient operator as follows:
\begin{equation} \label{grad-interm-supplemental}
\frac{\partial p(\{E_i\}_{i = 1}^{n}|H)}{\partial H_{\mu\nu}}  =
\sum_{m=1}^n {\rm tr} \bigg\{ E_{m} {\cal E}_{S+ER}(t_{m}) E_{m}
\otimes I_A \bigg[ \frac{\partial U(H)}{\partial H_{\mu\nu}} \,
\widetilde{\varrho}_{S+ER}(t_{m-1}) \otimes \varrho_A U^{\dag}(H)
+ U(H) \widetilde{\varrho}_{S+ER}(t_{m-1}) \otimes \varrho_A
\frac{\partial U^{\dag}(H)}{\partial H_{\mu\nu}} \bigg] \bigg\}.
\end{equation}

\noindent The operator $\frac{\partial p(\{E_i\}_{i =
1}^{n}|H)}{\partial H}$ is Hermitian provided the Hamiltonian $H$
is Hermitian. The derivative $\frac{\partial U(H)}{\partial H} =
\left. \frac{\partial U(H+V)}{\partial V} \right\vert_{V=0}$. We
use the perturbation expansion $\exp[-i(H+V)\tau] = \exp(-iH\tau)
T_{\leftarrow} \exp\left[ -i\int_0^\tau \exp(iHt')V\exp(-iHt') dt'
\right] = \exp(-iH\tau) - i \exp(-iH\tau) \int_0^\tau
\exp(iHt')V\exp(-iHt') dt' + o(V)$ and the spectral decomposition
$H=\sum_k \lambda_k |\psi_k\rangle \langle\psi_k|$ to get
\begin{equation} \label{U-derivative-supplemental}
\frac{\partial U(H)}{\partial H_{\mu\nu}} = \left( \frac{\partial
U^{\dag}(H)}{\partial H_{\mu\nu}} \right)^{\dag} = \sum_{k,l}
\frac{e^{-i\lambda_k\tau} - e^{-i\lambda_l \tau}}{\lambda_k -
\lambda_l} \langle \psi_k | \mu \rangle \langle \nu | \psi_l
\rangle \, |\psi_k\rangle \langle\psi_l|.
\end{equation}

Finally, we find the gradient of the logarithmic likelihood $\log
p({E_i}_{i=1}^n|H)$ with respect to the unknown parameters
$H_{\mu\nu}$:
\begin{equation} \label{log-gradient-supplemental}
\frac{\partial \log p(\{E_i\}_{i=1}^n|H)}{\partial H_{\mu\nu}} =
\frac{1}{p(\{E_i\}_{i=1}^n|H)} \, \frac{\partial
p(\{E_i\}_{i=1}^n|H)}{\partial H_{\mu\nu}}.
\end{equation}

\noindent Note that \eqref{log-gradient-supplemental} is
insensitive to the normalization of $p(\{E_i\}_{i=1}^n|H)$, which
significantly simplifies the calculation.

\begin{figure}[b]
\centering
\includegraphics[width=9cm]{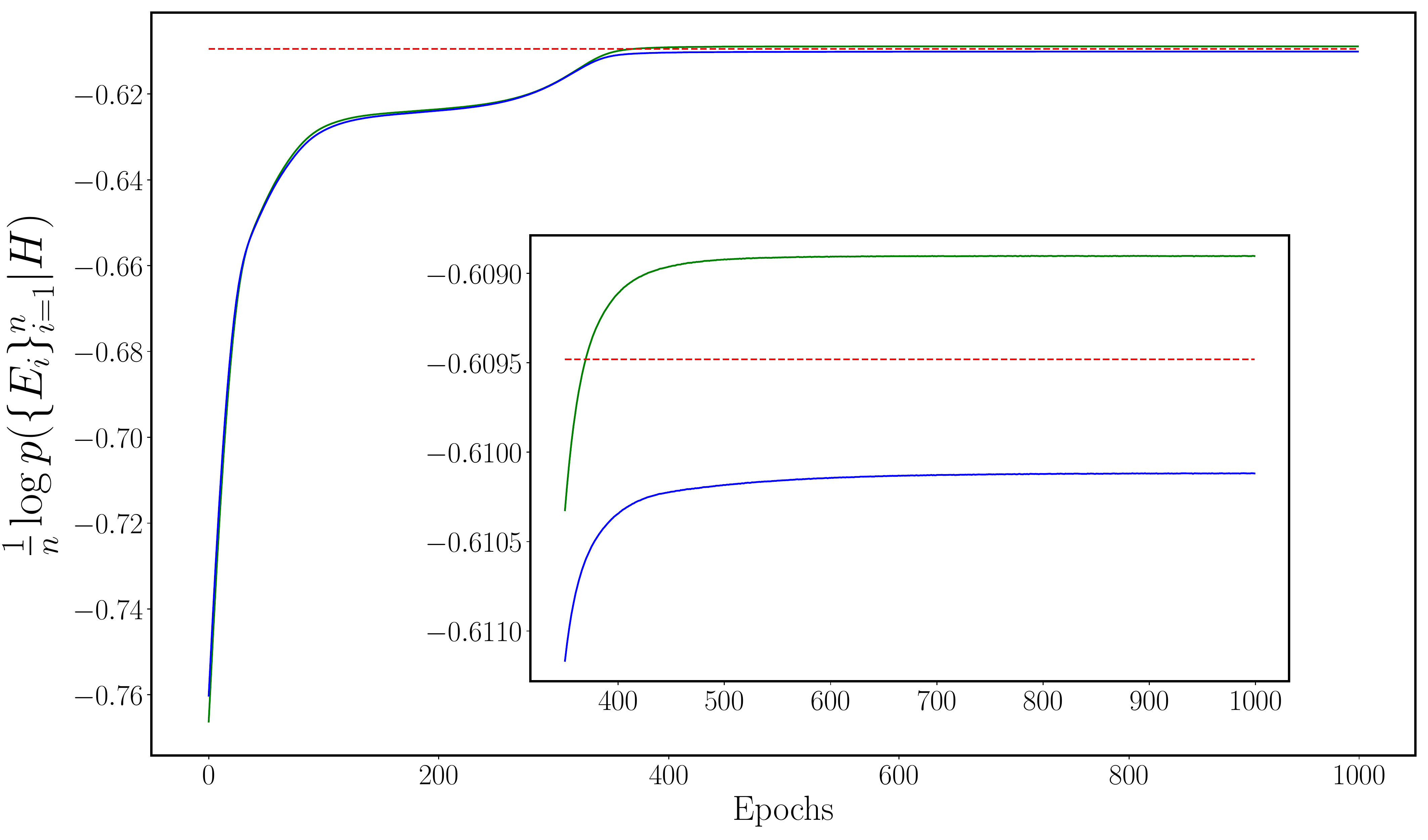}
\includegraphics[width=9cm]{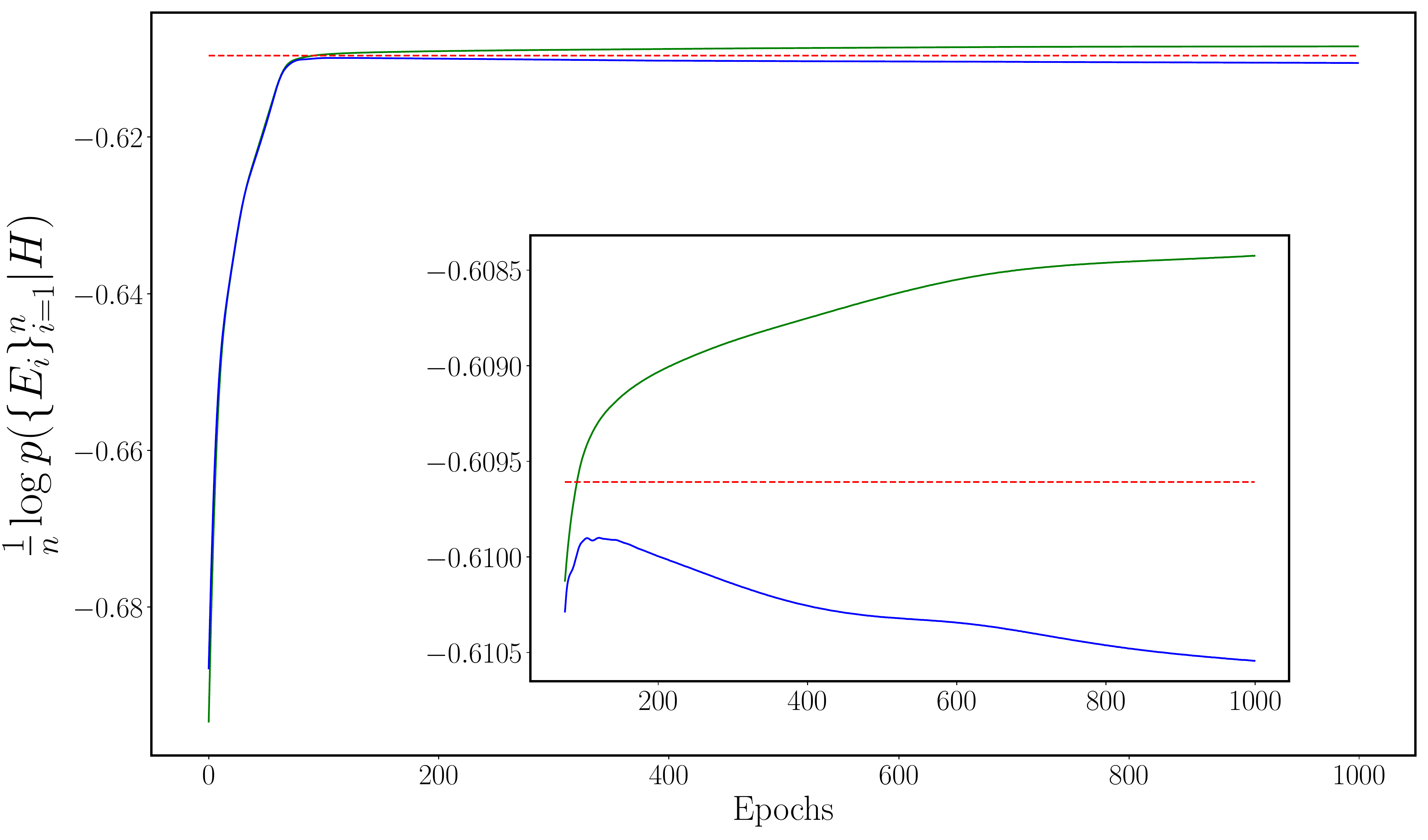}
\caption{Learning curves: The regularized logarithmic likelihood
for the training set (green, upper curve) and the regularized
logarithmic likelihood for the validation set (blue, lower curve)
vs. number of epochs in Adam algorithm~\cite{Adam}. The dimension
of the effective reservoir $d_{ER} = 2$ (left) and $d_{ER} = 6$
(right). Horizontal line is the theoretical value for the data
generated.} \label{figure-3-supplemental}
\end{figure}

\subsection{Details on the learning algorithm}

At item 1 of the algorithm, we initialize the model by randomly
choosing the factorized state $\varrho_{S+ER}(0)=\varrho_S(0)
\otimes \varrho_{ER}(0)$ and the factorized Hamiltonian
$H=H_{S+ER} \otimes I_{A}$. Starting with a Hamiltonian, which is
factorized with respect to $S+ER$ and $A$, fastens the learning
process of memory effects. Otherwise, the correlations between
$S+ER$ and $A$ induce irreducible decoherence and dissipation on
$S+ER$ that smear out the memory effects.

Typical learning curves in Fig.~\ref{figure-3-supplemental} show
how the logarithmic likelihood for the training set increases
during the learning process and approaches the theoretical
prediction, whereas the likelihood for the validation set starts
to decrease after some point in the case of overfitting ($d_{ER} =
6$). We test multiple variations of the batch size and the Adam
optimizer parameters~\cite{Adam} (used at step~6 of the learning
algorithm) to determine the fastest algorithm convergence. The
tuned parameters are $\beta_1=0.9$, $\beta_2=0.95$,
$\epsilon=10^{-4}$, the learning rate is $10^{-3}$, the batch size
is $10^3$.

\subsection{Details on overfitting}

The greater the dimension of the effective reservoir $d_{ER}$, the
greater the likelihood for the training set [see Fig.~4(a) in the
main text]. However, this leads to overfitting because the
likelihood for the validation set starts decreasing with the
growth of $d_{ER}$ above the optimal value. A discrepancy between
the likelihood for the training set and that for the validation
set is a direct indication of the bias-variance tradeoff in
machine learning~\cite{Bishop}. Fig.~\ref{figure-4-supplemental}
demonstrates the effect of overfitting on the quality of the
estimated open dynamics for the example studied. Non-optimal
hyperparameter $d_{ER} = 6$ leads to redundant oscillations in the
predicted dynamics (as opposed to the optimal value $d_{ER} = 2$).

\begin{figure}
\centering
\includegraphics[width=14cm]{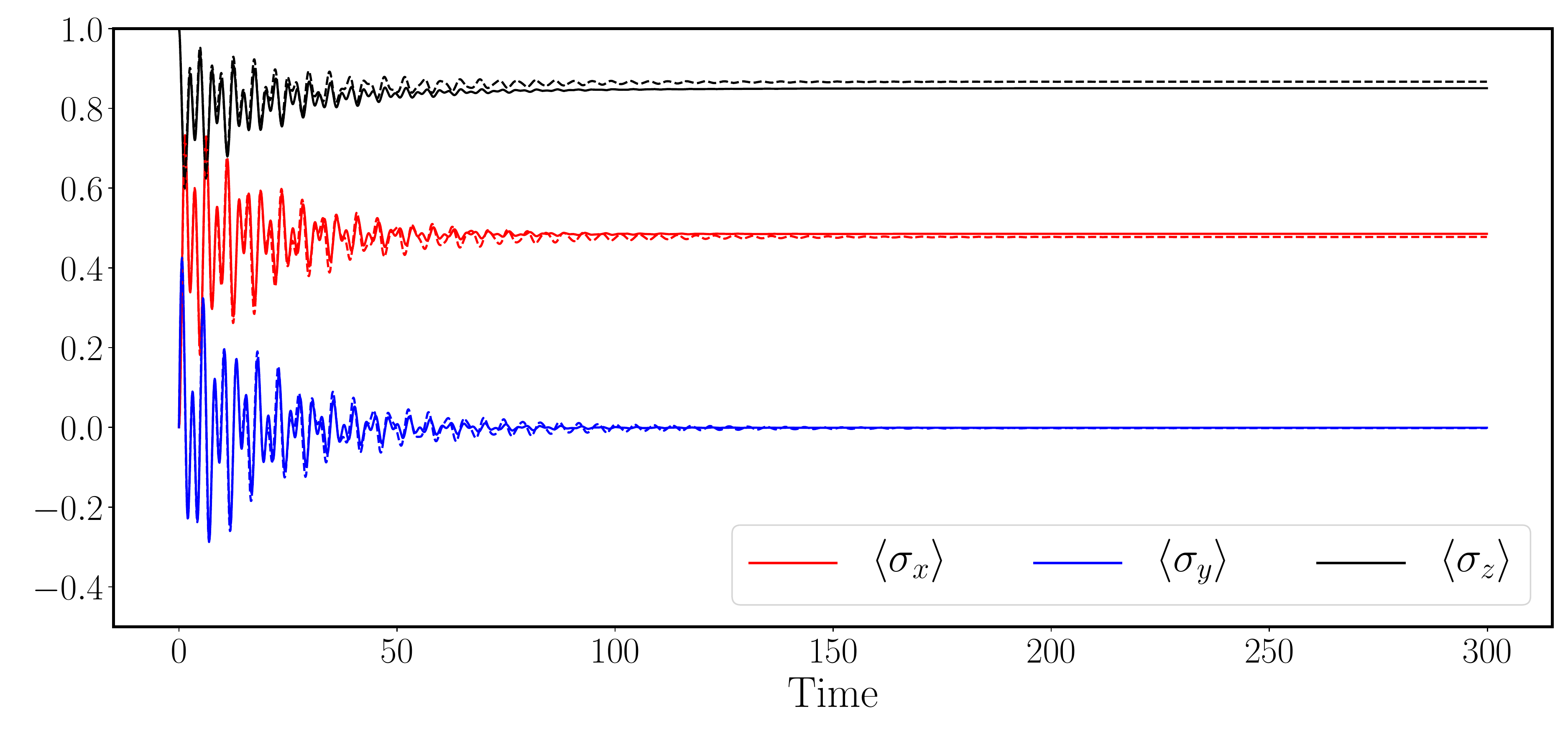}
\includegraphics[width=14cm]{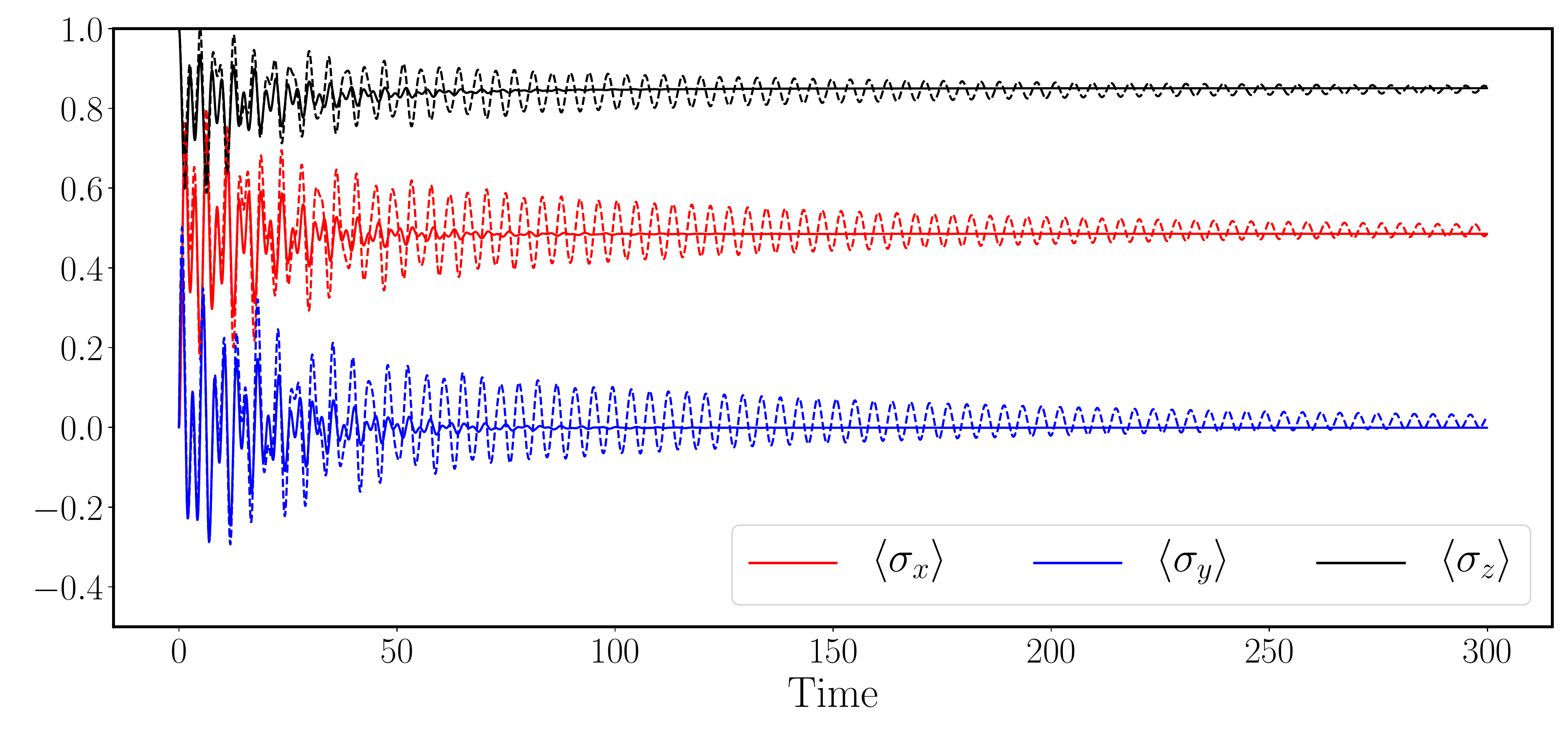}
\caption{Components $\langle \sigma_i(t) \rangle = {\rm
tr}[\varrho(t) \sigma_i]$, $i=x,y,z$, of the system Bloch vector
vs. dimensionless time for the exact open dynamics at a long time
scale (solid line) and the corresponding learning-based prediction
(dotted line) for $d_{ER} = 2$ (optimal, top panel) and $d_{ER} =
6$ (not optimal, bottom panel).} \label{figure-4-supplemental}
\end{figure}

Fig.~4(b) in the main text illustrates the perfect overfitting of
the learning algorithm if the dimension of the effective reservoir
$d_{ER} = d_S^n$, where $n$ is the number of projectors in the
data set $\{E_i\}_{i=1}^n$. Recall that $E_i = P_i \otimes
I_{ER}$, with $P_i =
\ket{\varphi_{k_i}^{(i)}}\bra{\varphi_{k_i}^{(i)}}$ being a pure
state of $S$. Consider the $n$-partite environment of the form
\begin{equation} \label{env-overfitting-supplemental}
\varrho_E = P_1 \otimes P_2 \otimes \ldots \otimes P_n
\end{equation}

\noindent and the time-independent unitary transformation for
$S+E$
\begin{eqnarray}
&& W = \exp(-iH\tau) = {\sf SHIFT} \cdot {\sf SWAP}, \\
&& {\sf SWAP} = \sum{i,j=1}^{d_S} \ket{i}_S\!\bra{j} \otimes
\ket{j}\bra{i} \otimes I_{2 \ldots n}, \\
&& {\sf SHIFT} = I_S \otimes \sum_{i_1,\ldots,i_n = 1}^{d_S}
\ket{i_2}\bra{i_1} \otimes \ket{i_1}\bra{i_2} \otimes
\ket{i_2}\bra{i_3} \otimes \ldots \otimes \ket{i_1}\bra{i_n}.
\end{eqnarray}

Note that $\varrho_S(\tau) = {\rm tr}_E (W \varrho_S(0) \otimes
\varrho_E W^{\dag})
 = P_1$ so the first projective measurement on the system in the
 basis $\ket{\varphi_{k_1}^{(1)}}$ would produce the specific
 outcome ($P_1$) with certainty, i.e., with probability 1. The
 state of environment after the first measurement on the system is
\begin{equation} \label{env-tau-overfitting-supplemental}
\varrho_E(\tau) = P_2 \otimes P_3 \otimes \ldots \otimes P_n
\otimes \varrho_S(0).
\end{equation}

\noindent Therefore, $\varrho_S(2\tau) = {\rm tr}_E (W
\varrho_S(\tau) \otimes \varrho_E(\tau) W^{\dag})
 = P_2$, i.e., the outcome for effect $P_2$ will be observed with
certainty. The state of environment after the first measurement on
the system is
\begin{equation} \label{env-2-tau-overfitting-supplemental}
\varrho_E(2\tau) = P_3 \otimes \ldots \otimes P_n \otimes
\varrho_S(0) \otimes P_1.
\end{equation}

One can continue the same line of reasoning until all $n$
measurements are performed. As a result we get
\begin{equation}
p(\{E_i\}_{i=1}^n) = \prod_{i=1}^n {\rm tr}[\varrho_S(i\tau) P_i]
= 1
\end{equation}

\noindent and the final state of the environment is
\begin{equation} \label{env-n-tau-overfitting-supplemental}
\varrho_E(n\tau) = \varrho_S(0) \otimes P_1 \ldots \otimes
P_{n-1}.
\end{equation}

\noindent This scenario corresponds to the perfect overfitting and
yields the logarithmic likelihood $\log p(\{E_i\}_{i=1}^n) = 0$
[depicted in the top right corner of Fig.~4(a) in the main text].

However, if more than $n$ measurements in random bases are
performed, then the proposed effective reservoir of dimension
$d_S^n$ fails in reproducing the results perfectly. In fact,
continue the process above with an extended series of measurements
$\{E_i\}_{i=n+1}^{2n}$, then
\begin{equation}
p(\{E_i\}_{i=n+1}^{2n}) = {\rm tr}[\varrho_S(0) P_{n+1}]
\prod_{i=1}^{n-1} {\rm tr}[P_{i} P_{n+i+1}].
\end{equation}

\noindent If the measurement bases are chosen randomly (Haar
measure), then the average $\braket{{\rm tr}[P_{i} P_{n+i+1}]} =
\frac{1}{d_S}$. Concavity of the logarithm implies
\begin{equation}
\lim_{n \rightarrow \infty} \frac{1}{n} \log
p(\{E_i\}_{i=n+1}^{2n}) \leq \log\left[ \lim_{n \rightarrow
\infty} \frac{1}{n} \left( {\rm tr}[\varrho_S(0) P_{n+1}] +
\sum_{i=1}^{n-1} {\rm tr}[P_{i} P_{n+i+1}] \right) \right] = \log
\frac{1}{d_S}.
\end{equation}

\noindent In other words, the regularized logarithmic likelihood
on the validation set $\{E_i\}_{i=n+1}^{2n}$ tends to a value not
exceeding $\log \frac{1}{d_S}$, which is approximately $-0.69$ for
$d_S = 2$ [see the bottom right corner of Fig.~4(a) in the main
text].

\subsection{Details on the hyperparameter $d_{ER}$}

In general, the hyperparameter $d_{ER}$ is tuned in such a way
that the likelihood on the validation set achieves its maximum.
Tuning is reasonable to perform in the vicinity of the physical
estimate~\cite{luchnikov-2019}
\begin{equation} \label{der-sufficient}
d_{ER}(\epsilon) = \min_{0 < \alpha < 1}
\frac{\sqrt{1-\alpha}}{\epsilon^{\alpha/2(1-\alpha)}} \exp \left[
n \gamma T \, \frac{(\gamma \tau)^{\alpha-1}-\alpha}{1-\alpha}
\right],
\end{equation}

\noindent where $\epsilon$ is the desired accuracy of open
dynamics, $n$ is the effective subsystem's number of degrees of
freedom interacting with the reservoir ($n \leq d_S^2$), $\gamma$
is the coupling strength (between the open system and the
reservoir), $T$ is the reservoir correlation time, and $\tau$ is
the minimal timescale for the open system dynamics (the inverse of
the cutoff frequency for the reservoir spectral function).

For some physical systems $d_{ER}$ is defined by the very
reservoir structure. For instance, the nuclear spin $I = 1$ in
nitrogen ${}^{14}N$ is as an effective reservoir for the
electronic spin qubit in a nitrogen-vacancy center in
diamond~\cite{haase-2018}, which implies $d_{ER} = 2I+1 = 3$.

\subsection{Details on the data generation}

The synthetic training set $\{E_i\}_{i=1}^n$ is generated in a
non-Markovian composite bipartite collision
model~\cite{lorenzo-2017}. We consider a bipartite system that
successively interacts with qubit subenvironments $R$ during
collision time $\Delta t$, with the bipartite system being
composed of the very open qubit system under study $S$ and one
auxiliary qubit system $S_1$, see
Fig.~\ref{figure-5-supplemental}. The composite bipartite
collision model~\cite{lorenzo-2017} allows to find the system
evolution $\varrho_S(t) = {\rm tr}_{S_1}[\varrho_{S+S_1}(t)]$
intervened by measurements on the system.

\begin{figure}
\centering
\includegraphics[width=12cm]{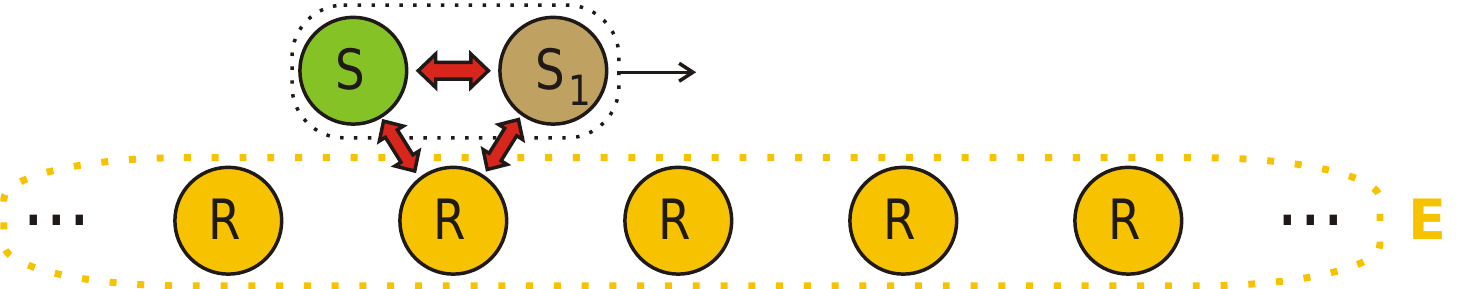}
\caption{Composite bipartite collision model used in data
generation. The infinite environment $E$ is composed of identical
qubit subenvironments $R$ each with a density matrix $\varrho_R =
\frac{1}{2}(I + \sigma_z)$. Three thick red arrows denote the
interaction with
Hamiltonian~\eqref{Hamiltonian-collision-supplemental} for a
period $\Delta t$.} \label{figure-5-supplemental}
\end{figure}

We fix the (dimensionless) interaction Hamiltonian between $S$,
$S_1$, and $R$ in the form
\begin{equation} \label{Hamiltonian-collision-supplemental}
H_{S+S_1+R} = \sigma_z \otimes I \otimes I +  \sigma_x \otimes I
\otimes I + I \otimes  \sigma_z \otimes I +  I \otimes  \sigma_x
\otimes I +\sigma_z \otimes \sigma_z \otimes I + 0.3 \, I \otimes
\sigma_z \otimes \sigma_z +  0.3 \, I \otimes \sigma_y \otimes
\sigma_y +0.3 \, I \otimes \sigma_x \otimes \sigma_x.
\end{equation}

\noindent The coefficients in the interaction Hamiltonian
correspond to the case when strong memory effects are present in
the evolution whereas the relaxation time is much longer than the
recurrence time of memory effects --- the hardest open dynamics to
reconstruct. Each collision results in the transformation
$\varrho_{S+S_1}(t+\Delta t) = {\rm tr}_{R} \big[ \exp(-i
H_{S+S_1+R} \Delta t) \varrho_{S+S_1}(t) \otimes \varrho_{R}
\exp(i H_{S+S_1+R} \Delta t) \big]$.

To simulate projective measurements in random bases we proceed as
follows. Suppose the qubit system is in the state $\varrho_S(t_i)$
at time $t_i = i \tau$. We randomly choose a direction ${\bf
r}^{(i)}\in \mathbb{R}^3$, $|{\bf r}^{(i)}|=1$, on a Bloch ball
and calculate eigenvectors $|\varphi_{+}^{(i)}\rangle$ and
$|\varphi_{-}^{(i)}\rangle$ of the polarization operator
$r_x^{(i)} \sigma_x + r_y^{(i)} \sigma_y + r_z^{(i)} \sigma_z$,
where $(\sigma_x,\sigma_y,\sigma_z)$ is the conventional set of
Pauli operators. The transformation $\{\pm\} \rightarrow
|\varphi_{\pm}^{(i)}\rangle \langle\varphi_{\pm}^{(i)}|$ is an
observable at time $t_i = i \tau$. One of the two measurement
outcomes $\{\pm\}$ is accepted, with the probability to accept the
result $+$ being $\langle\varphi_{+}^{(i)}| \varrho_S(t)
|\varphi_{+}^{(i)}\rangle$. As a result, one of the operators
$|\varphi_{\pm}^{(i)}\rangle \langle\varphi_{\pm}^{(i)}| \otimes
I_{ER}$ is accepted as $E_i$. Observation of the outcome $\pm$ in
the $i$th measurement of the system results in the transformation
$\varrho_{S+S_1} \rightarrow |\varphi_{\pm}^{(i)}\rangle
\langle\varphi_{\pm}^{(i)}| \otimes \varrho_{S_1}^{\pm} / {\rm
tr}[\varrho_{S_1}^{\pm}]$, where $\varrho_{S_1}^{\pm} =
(\langle\varphi_{\pm}^{(i)}| \otimes I_{S_1}) \varrho_{S+S_1}(t)
(|\varphi_{\pm}^{(i)}\rangle \otimes I_{S_1})$. The measurement is
followed by another collision described above, which in turn is
followed by a measurement, and so on until the set
$\{E_i\}_{i=1}^n$ is completed.

\subsection{Variational Bayesian inference approach}

In the presented learning algorithm, we maximize the likelihood
function and find parameters $H_{\mu\nu}$ encoding the desired
generator ${\cal L}_{S+ER}$ for the Markovian embedding. However,
the algorithm itself does not provide the error (variance) of
parameters $H_{\mu\nu}$. This error can, however, be estimated via
the variational Bayesian method as follows. Let $p(H)$ be an
\textit{a priori} distribution for Hamiltonian $H$, say, a
\textit{uniform} distribution on ${\rm Re}H_{\mu\nu}$ and ${\rm
Im}H_{\mu\nu}$ within a wide range. For an observed sequence of
operators $\{E_i\}_{i=1}^n$ we get the \textit{a posteriori}
distribution
\begin{equation}
p(H|\{E_i\}_{i=1}^n) = \frac{p(\{E_i\}_{i=1}^n|H) p(H)}{\int
p(\{E_i\}_{i=1}^n|H) p(H) dH} = \frac{1}{Z} p(\{E_i\}_{i=1}^n|H)
\end{equation}

\noindent with some constant $Z$. Although $p(H|\{E_i\}_{i=1}^n)$
is not known precisely, we expect that for sufficiently big data
set this distribution can be approximated by a factorized Gaussian
distribution
\begin{equation}
Q_{\{\kappa_{\mu\nu},\sigma_{\mu\nu},\varkappa_{\mu\nu},\varsigma_{\mu\nu}\}}(H)
= \prod_{\mu,\nu} \left[ \frac{1}{\sqrt{2\pi} \, \sigma_{\mu\nu}}
\exp \left( - \frac{({\rm Re}H_{\mu\nu} -
\kappa_{\mu\nu})^2}{2\sigma_{\mu\nu}^2} \right) \right]
\prod_{\mu,\nu} \left[ \frac{1}{\sqrt{2\pi} \, \varsigma_{\mu\nu}}
\exp \left( - \frac{({\rm Im}H_{\mu\nu} -
\varkappa_{\mu\nu})^2}{2\varsigma_{\mu\nu}^2} \right) \right],
\end{equation}

\noindent where the parameters $\kappa_{\mu\nu} + i
\varkappa_{\mu\nu}$ define the optimal values $H_{\mu\nu}$
maximizing the likelihood, and the standard deviations
$\sigma_{\mu\nu}$ and $\varsigma_{\mu\nu}$ define the accuracy of
parameter estimation for the real and imaginary part of
$H_{\mu\nu}$, respectively. If the number of measurements $n
\rightarrow \infty$, then $\sigma_{\mu\nu} \rightarrow 0$ and
$\varsigma_{\mu\nu} \rightarrow 0$. Our goal is to find
$\sigma_{\mu\nu}$ and $\varsigma_{\mu\nu}$ for a finite $n$. To do
so we minimize the Kullback--Leibler divergence $D_{\rm KL}$ of
$p(H|\{E_i\}_{i=1}^n)$ from
$Q_{\{\kappa_{\mu\nu},\sigma_{\mu\nu},\varkappa_{\mu\nu},\varsigma_{\mu\nu}\}}(H)$,
where $D_{\rm KL}(q(x) \, || \, p(x)) = \int dx \, q(x) [\log q(x)
- \log p(x)]$. One can readily see that
\begin{equation} \label{KL-relation}
D_{\rm KL} \left(
Q_{\{\kappa_{\mu\nu},\sigma_{\mu\nu},\varkappa_{\mu\nu},\varsigma_{\mu\nu}\}}(H)
\, || \, p(H|\{E_i\}_{i=1}^n) \right) = D_{\rm KL} \left(
Q_{\{\kappa_{\mu\nu},\sigma_{\mu\nu},\varkappa_{\mu\nu},\varsigma_{\mu\nu}\}}(H)
\, || \, p(\{E_i\}_{i=1}^n |H) \right) + \log Z.
\end{equation}

\noindent Since $Z$ is independent of $H$ or either of
$\kappa_{\mu\nu},\sigma_{\mu\nu},\varkappa_{\mu\nu},\varsigma_{\mu\nu}$,
the minimization of~\eqref{KL-relation} reduces to the
minimization of $D_{\rm KL} \left(
Q_{\{\kappa_{\mu\nu},\sigma_{\mu\nu},\varkappa_{\mu\nu},\varsigma_{\mu\nu}\}}(H)
\, || \, p(\{E_i\}_{i=1}^n |H) \right)$. Using the explicit form
of the Gaussian distribution, we come the so-called
reparameterization trick~\cite{molchanov-2017,kingma-2013}: the
minimization of $D_{\rm KL} \left(
Q_{\{\kappa_{\mu\nu},\sigma_{\mu\nu},\varkappa_{\mu\nu},\varsigma_{\mu\nu}\}}(H)
\, || \, p(\{E_i\}_{i=1}^n |H) \right)$ is equivalent to the
minimization of the functional
\begin{equation} \label{F-supplemental}
F(\kappa_{\mu\nu},\sigma_{\mu\nu},\varkappa_{\mu\nu},\varsigma_{\mu\nu})
= - \sum_{\mu\nu} \log \sigma_{\mu\nu} - \sum_{\mu\nu} \log
\varsigma_{\mu\nu} - \mathbb{E}_{\scriptsize \begin{array}{c}
  \xi_{\mu\nu} \sim {\cal N}(0,1) \\
  \zeta_{\mu\nu} \sim {\cal N}(0,1) \\
\end{array}} \left[ p\Big( \{E_i\}_{i=1}^n \Big| \{ \kappa_{\mu\nu}+
\xi_{\mu\nu} \sigma_{\mu\nu} + i(\varkappa_{\mu\nu}+
\zeta_{\mu\nu} \varsigma_{\mu\nu}) \}_{\mu,\nu} \Big) \right],
\end{equation}

\noindent where ${\cal N}(0,1)$ is the standard normal
distribution. The expectation value in the right hand side of
Eq.~\eqref{F-supplemental} is readily estimated by sampling from
the standard normal distribution, i.e., $\mathbb{E}_{\xi \sim
{\cal N}(0,1)} f(\xi) \approx \frac{1}{M} \sum_{j=1}^M
f(\xi^{(j)})$, where $\xi^{(j)}$ is a sample from ${\cal N}(0,1)$.
With such an estimation at hand, the minimization
of~\eqref{F-supplemental} is readily performed numerically, which
yields the optimal approximation $Q_{\{\kappa_{\mu\nu}^{\rm
opt},\sigma_{\mu\nu}^{\rm opt},\varkappa_{\mu\nu}^{\rm
opt},\varsigma_{\mu\nu}^{\rm opt}\}}(H)$.

\textit{A posteriori} distribution of the system density operator
$\varrho_S(t)$ at time $t$ is
\begin{eqnarray} \label{rho-sampling-supplemental}
p\big( \varrho_S(t) \big| \{E_i\}_{i=1}^n \big) &=& \int
\delta\left\{ \varrho_{S}(t) - {\rm tr}_{ER} \left[ \exp\big(t
{\cal L}_{S+ER}(H) \big) \varrho_{S+ER}(0) \right] \right\}
p(H|\{E_i\}_{i=1}^n) \, dH \nonumber\\
& \approx & \int \delta\left\{ \varrho_{S}(t) - {\rm tr}_{ER}
\left[ \exp\big(t {\cal L}_{S+ER}(H) \big) \varrho_{S+ER}(0)
\right] \right\} Q_{\{\kappa_{\mu\nu}^{\rm
opt},\sigma_{\mu\nu}^{\rm opt},\varkappa_{\mu\nu}^{\rm
opt},\varsigma_{\mu\nu}^{\rm opt}\}}(H) \, dH.
\end{eqnarray}

\noindent Sampling from the
distribution~\eqref{rho-sampling-supplemental} for a given time
$t$, we get the standard deviation for matrix elements of the
system density operator $\varrho_S(t)$. The results are depicted
in Fig.~\ref{figure-6-supplemental} for short and long timescales.
The maximum standard deviation for matrix elements of the
estimated density operator equals 0.025 at time moments $t=50$.

\begin{figure}
\centering
\includegraphics[width=14cm]{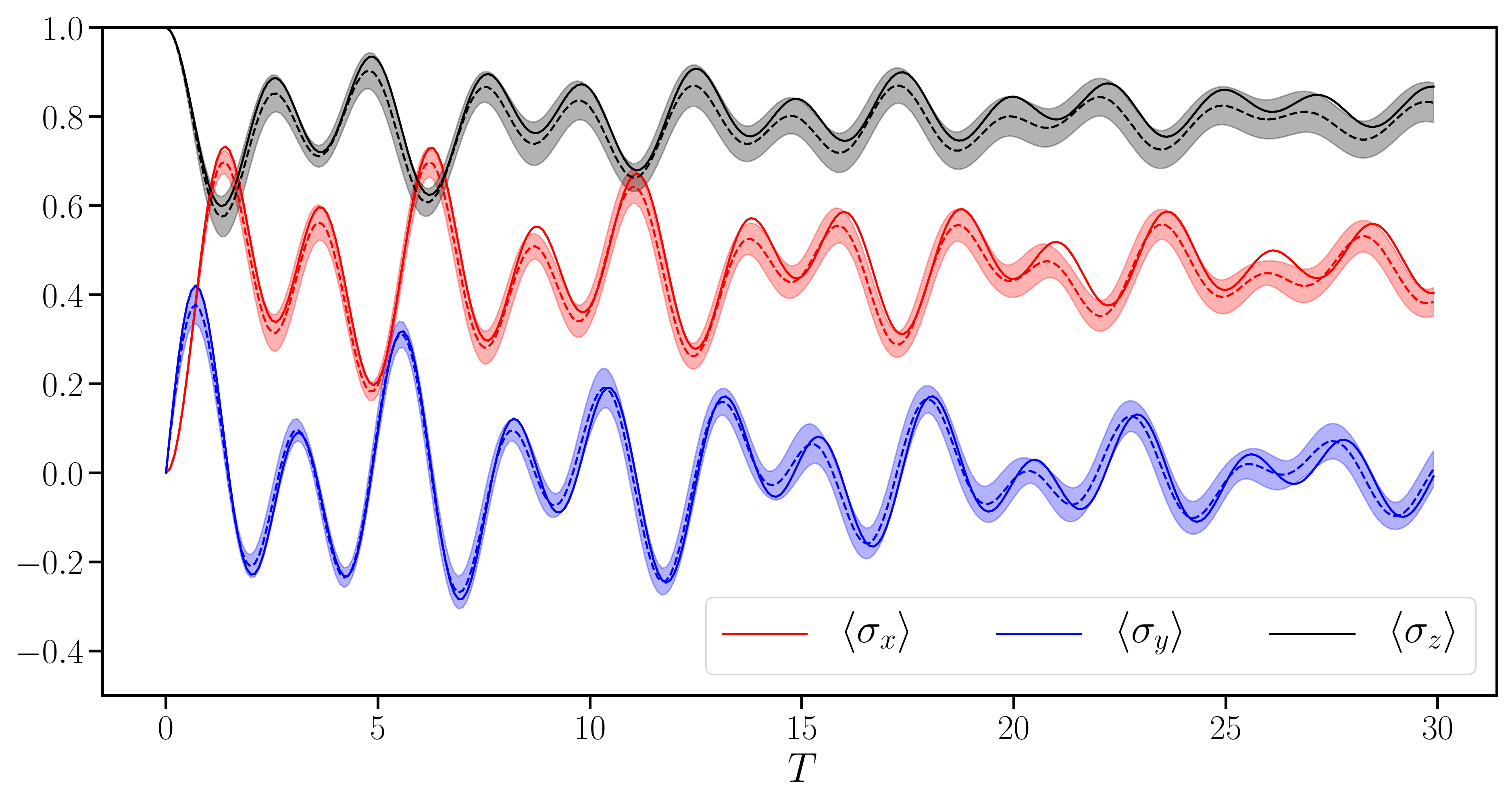}
\includegraphics[width=14cm]{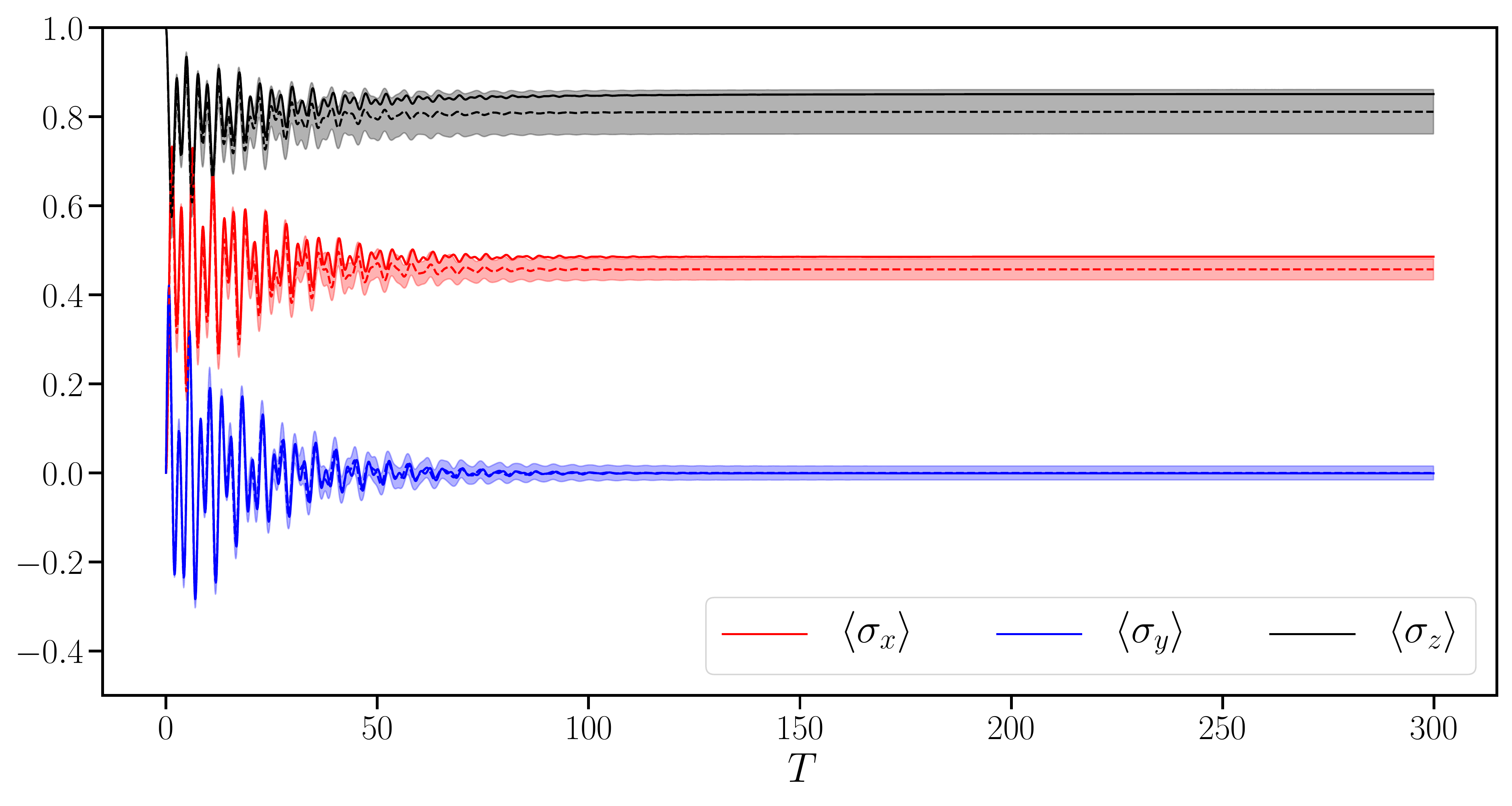}
\caption{Exact solution (solid line) and the Bayesian inference
(dotted line, $d_{ER}=2$) for the open qubit dynamics in terms of
the Bloch vector at a short timescale (top panel) and a long time
scale (bottom panel). Shaded area depicts the standard deviation
for the Bloch vector components.} \label{figure-6-supplemental}
\end{figure}

\subsection{Error estimation}

\begin{figure}[b]
\centering
\includegraphics[width=12cm]{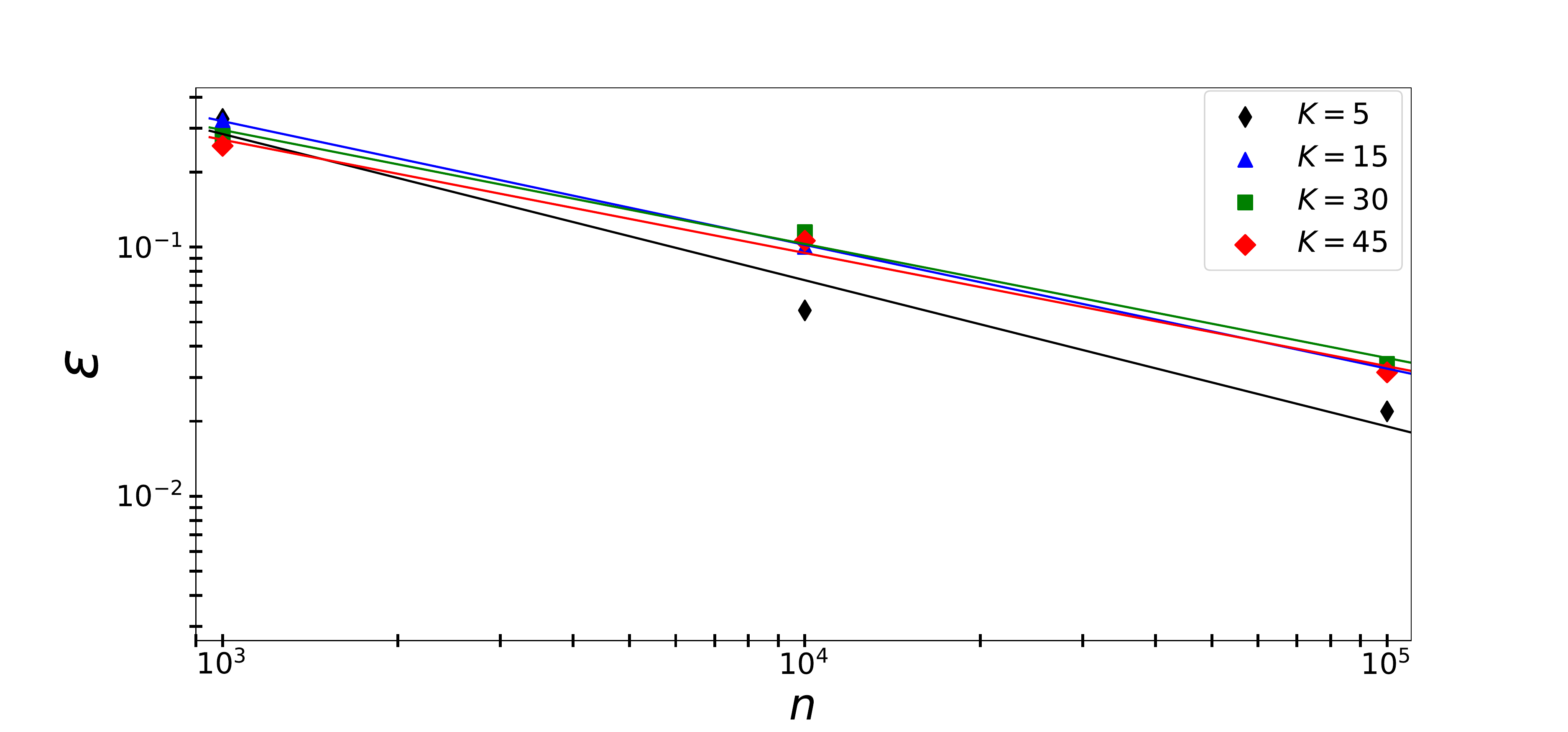}
\includegraphics[width=12cm]{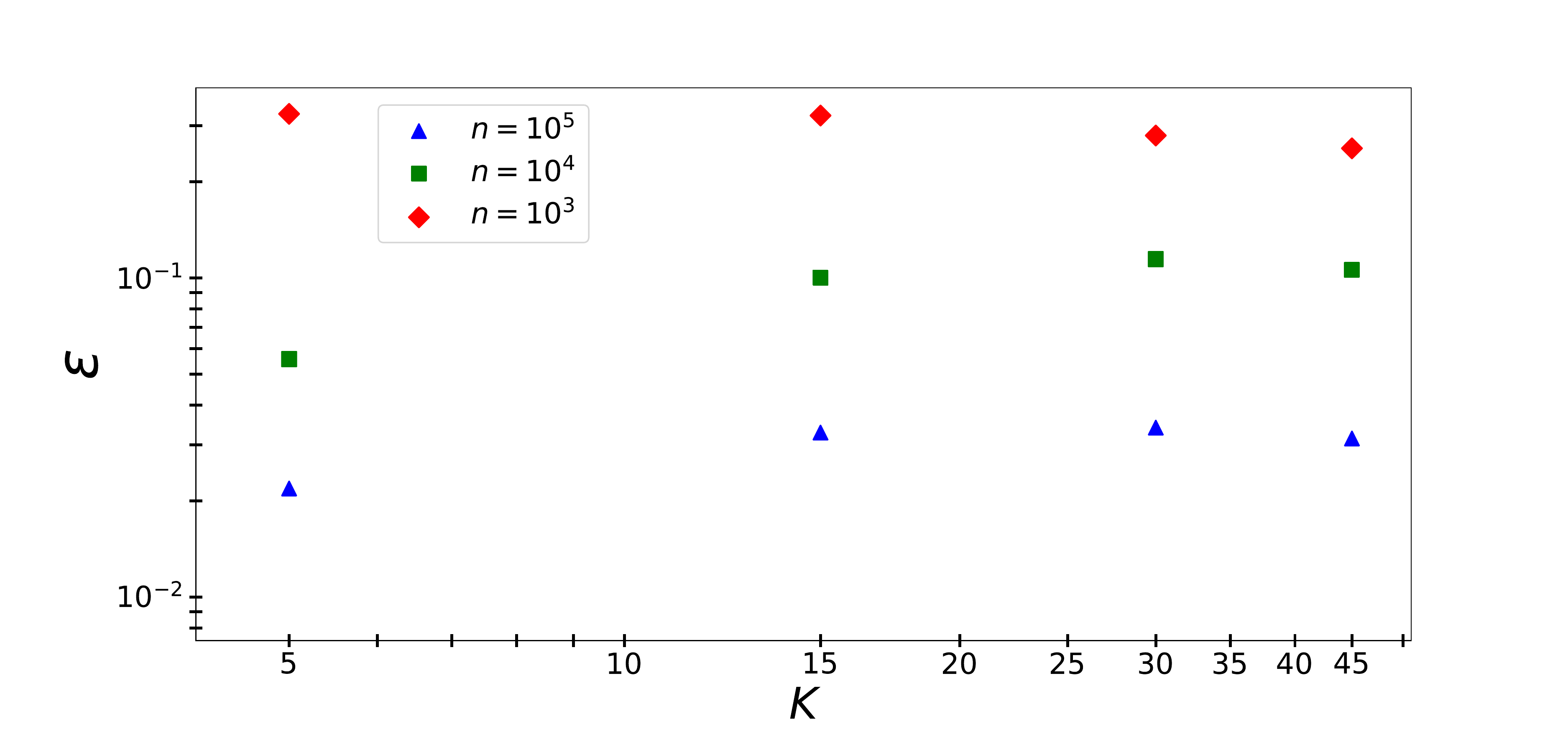}
\caption{Average error~\eqref{error-average-supplemental} for the
estimated quantum dynamical map $\Phi_S(t)$ with $d_{ER}=2$ vs.
the number of projective measurements used (top panel), vs. the
number $K$ of time moments analyzed, $t_i = i$, $i=1, \ldots, K$
(bottom panel).} \label{figure-7-supplemental}
\end{figure}

\begin{figure}
\centering
\includegraphics[width=12cm]{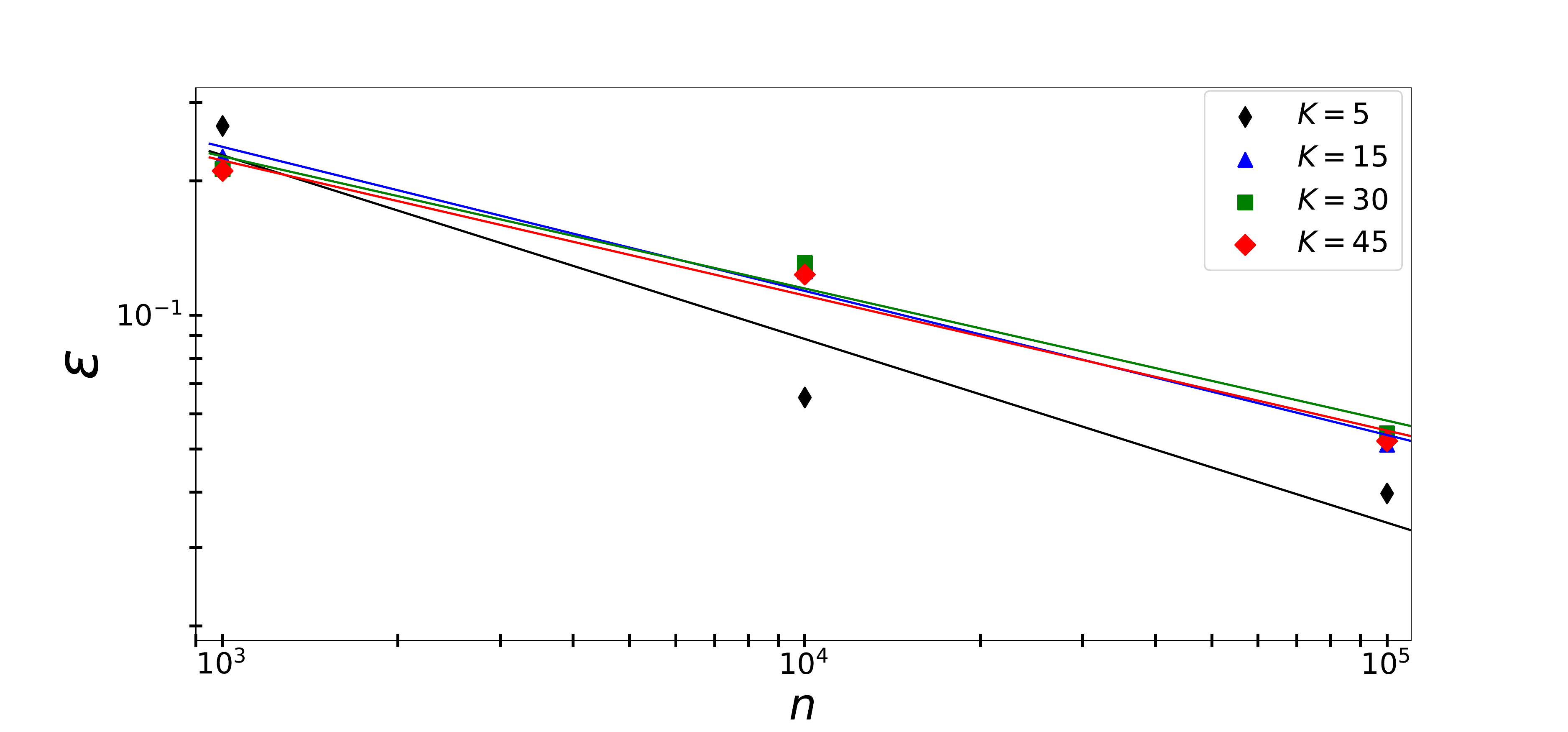}
\includegraphics[width=12cm]{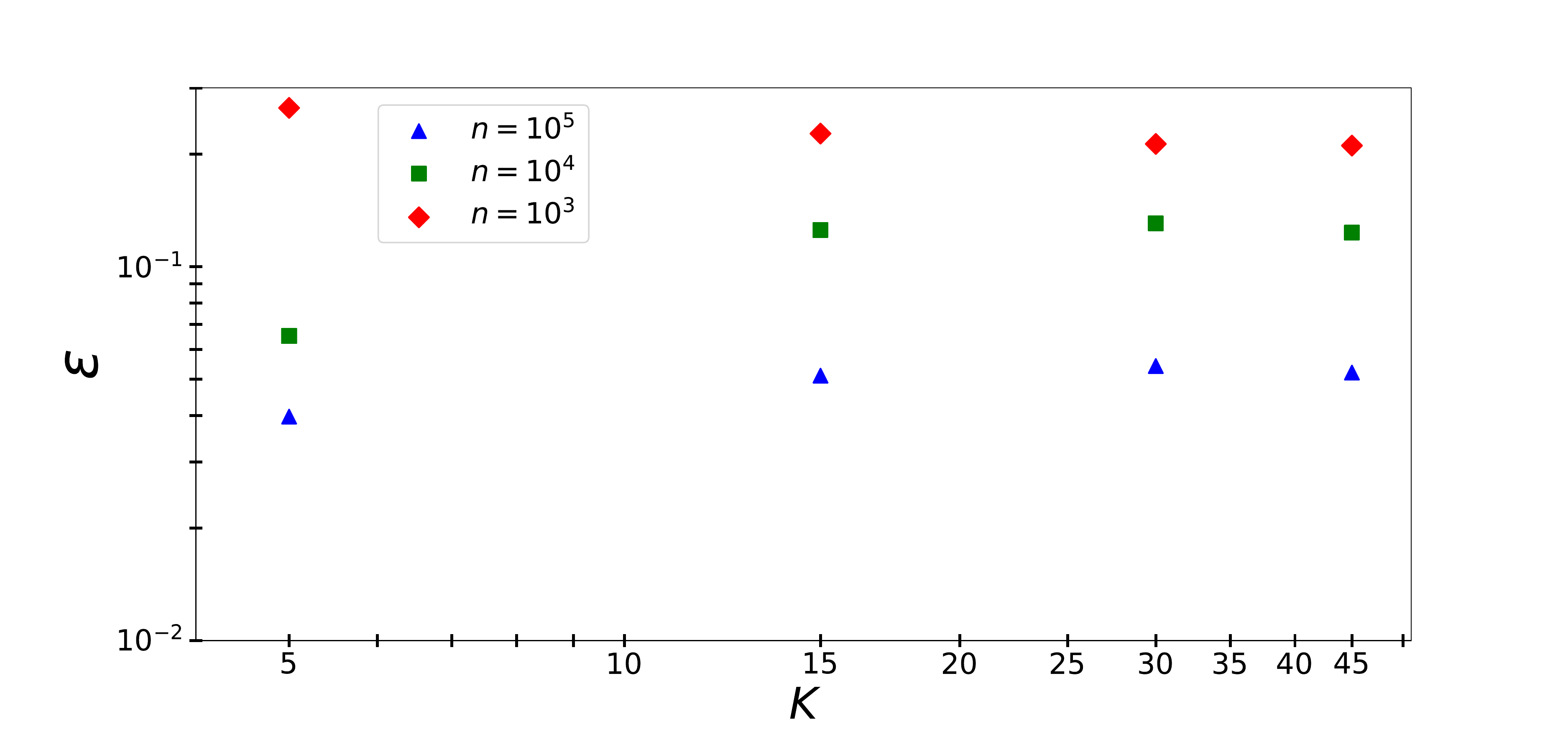}
\caption{Average error~\eqref{error-average-B-supplemental} for
the Bayesian inference for the quantum dynamical map $\Phi_S(t)$
with $d_{ER}=2$ vs. the number of projective measurements used
(top panel), vs. the number $K$ of time moments analyzed, $t_i =
i$, $i=1, \ldots, K$ (bottom panel).}
\label{figure-8-supplemental}
\end{figure}

The presented learning algorithm uses $n$ projective measurements
on the system to estimate the Markovian embedding for the
non-Markovian system dynamics. This reconstruction results in the
process tensor depicted in Fig.~2 in the main text. As a result,
we can infer a desired number of channels $\Phi_S(t_i)$, $i=1,
\ldots, K$, for the system dynamics from time $0$ to time $t_i$ by
formula
\begin{equation}
\Phi_S(t)[\varrho_S(0)] = {\rm tr}_{ER} \left[ \exp(t {\cal
L}_{S+ER}) [\varrho_S(0) \otimes \varrho_{ER}(0)] \right].
\end{equation}

By the Choi--Jamio{\l}kowski isomorphism, all the information
about the channel $\Phi_S(t)$ is contained in the matrix
$\Omega_{\Phi_S(t)}$ defined through (see, e.g.,
\cite{Holevo,filippov-jms-2019})
\begin{equation}
\Omega_{\Phi_S(t)} = (\Phi_S(t) \otimes {\rm Id}_S)
[\ket{\psi_+}\bra{\psi_+}],
\end{equation}

\noindent where $\ket{\psi_+} = \frac{1}{\sqrt{d_S}}
\sum_{i=1}^{d_S} \ket{i} \otimes \ket{i}$ is the maximally
entangled state.

Provided the exact dynamical map $\Phi_S^{\rm exact}(t)$ is known,
the error in estimating the channel $\Phi_S(t)$ can therefore be
expressed as
\begin{equation}
\epsilon\big(\Phi_S(t)\big) = \frac{1}{2} \| \Omega_{\Phi_S(t)} -
\Omega_{\Phi_S^{\rm exact}(t)} \|_1,
\end{equation}

\noindent where $\|A\|_1 = {\rm tr}\sqrt{A^{\dag}A}$. The average
error in estimating a set of channels $\{\Phi_S(t_i)\}_{i=1}^K$
equals
\begin{equation} \label{error-average-supplemental}
\varepsilon(\{\Phi_S(t_i)\}_{i=1}^K) = \frac{1}{K} \sum_{i=1}^{K}
\epsilon(\Phi_S(t_i)) = \frac{1}{2K} \sum_{i=1}^{K} \|
\Omega_{\Phi_S(t_i)} - \Omega_{\Phi_S^{\rm exact}(t_i)} \|_1.
\end{equation}

Fig.~\ref{figure-7-supplemental} shows the error
$\varepsilon(\{\Phi_S(t_i)\}_{i=1}^K)$ scales with $n$ as
$\frac{1}{\sqrt{n}}$ and is almost independent of $K$.

If the exact quantum dynamical map $\Phi_S(t)$ is not known (as it
takes place for experimental data), the error is estimated via the
variational Bayesian inference approach. In full analogy with the
previous section, we get the posterior distribution of channels
$\Phi_S(t)$ and the corresponding Choi operators, namely,
\begin{equation}
p( \Omega_{\Phi_S(t)} | \{E_i\}_{i=1}^n ) \approx \int
\delta\left\{ \Omega_{\Phi_S(t)} - \frac{1}{d_S}
\sum_{i,j=1}^{d_S} {\rm tr}_{ER} \left[ \exp(t {\cal L}_{S+ER}) [
\ket{i}\bra{j} \otimes \varrho_{ER}(0)] \right] \otimes
\ket{i}\bra{j} \right\} Q_{\{\kappa_{\mu\nu}^{\rm
opt},\sigma_{\mu\nu}^{\rm opt},\varkappa_{\mu\nu}^{\rm
opt},\varsigma_{\mu\nu}^{\rm opt}\}}(H) \, dH.
\end{equation}

\noindent By sampling from the latter distribution, we calculate
the average error for the proposed learning algorithm
\begin{equation} \label{error-average-B-supplemental}
\varepsilon(\{\Phi_S(t_i)\}_{i=1}^K) = \frac{1}{2K} \sum_{i=1}^{K}
\mathbb{E}\left[ \| \Omega_{\Phi_S(t_i)} -
\Omega_{\mathbb{E}\Phi_S(t_i)} \|_1 \right],
\end{equation}

\noindent where $\mathbb{E}\Phi_S(t_i)$ is the mean Bayesian
inference (obtained via averaging over samples).
Fig.~\ref{figure-8-supplemental} shows that, in this case, the
error $\varepsilon(\{\Phi_S(t_i)\}_{i=1}^K)$ scales with $n$ as
$\frac{1}{\sqrt{n}}$ and is almost independent of $K$.

\subsection{Comparison with the full process tomography}

\begin{figure}
\centering
\includegraphics[width=12cm]{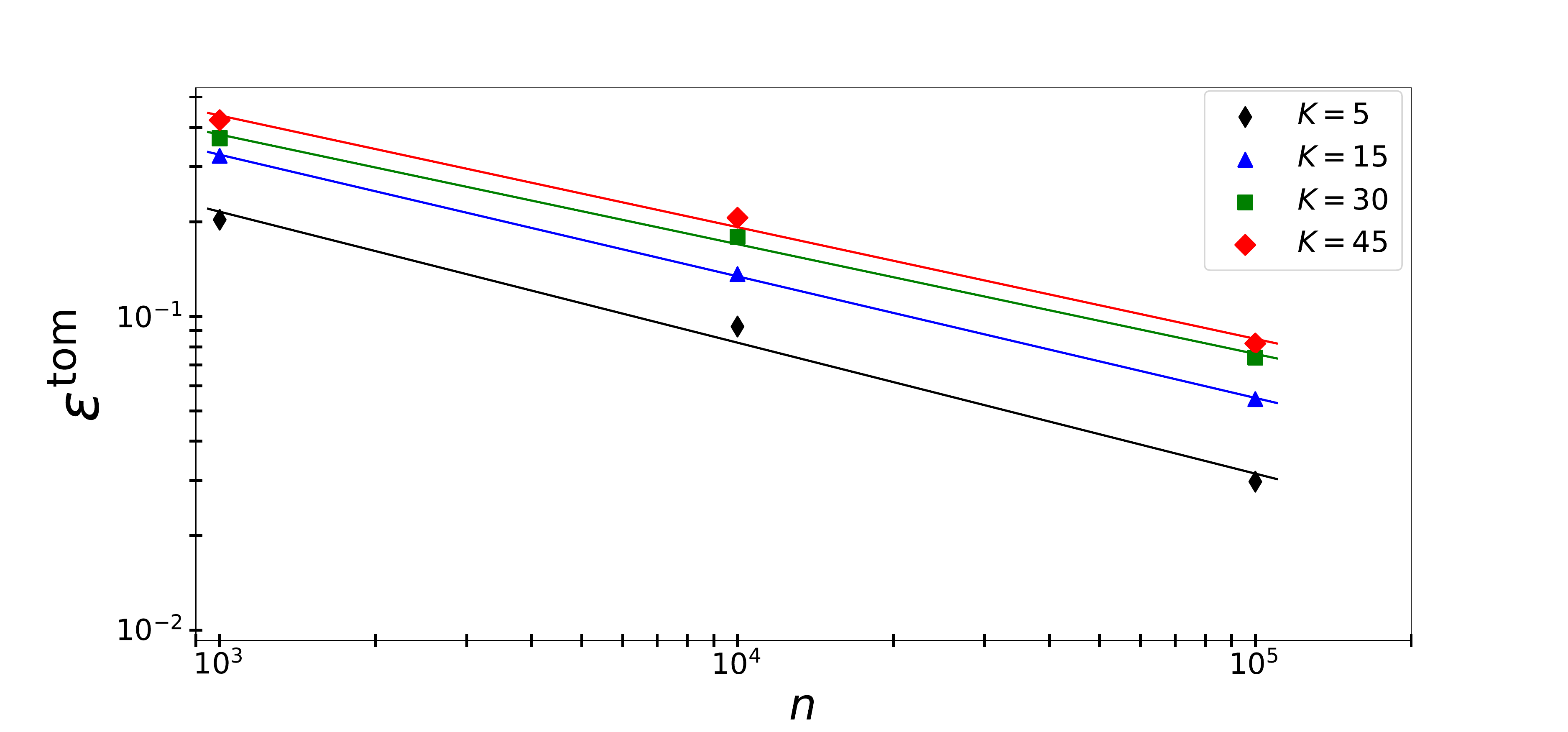}
\includegraphics[width=12cm]{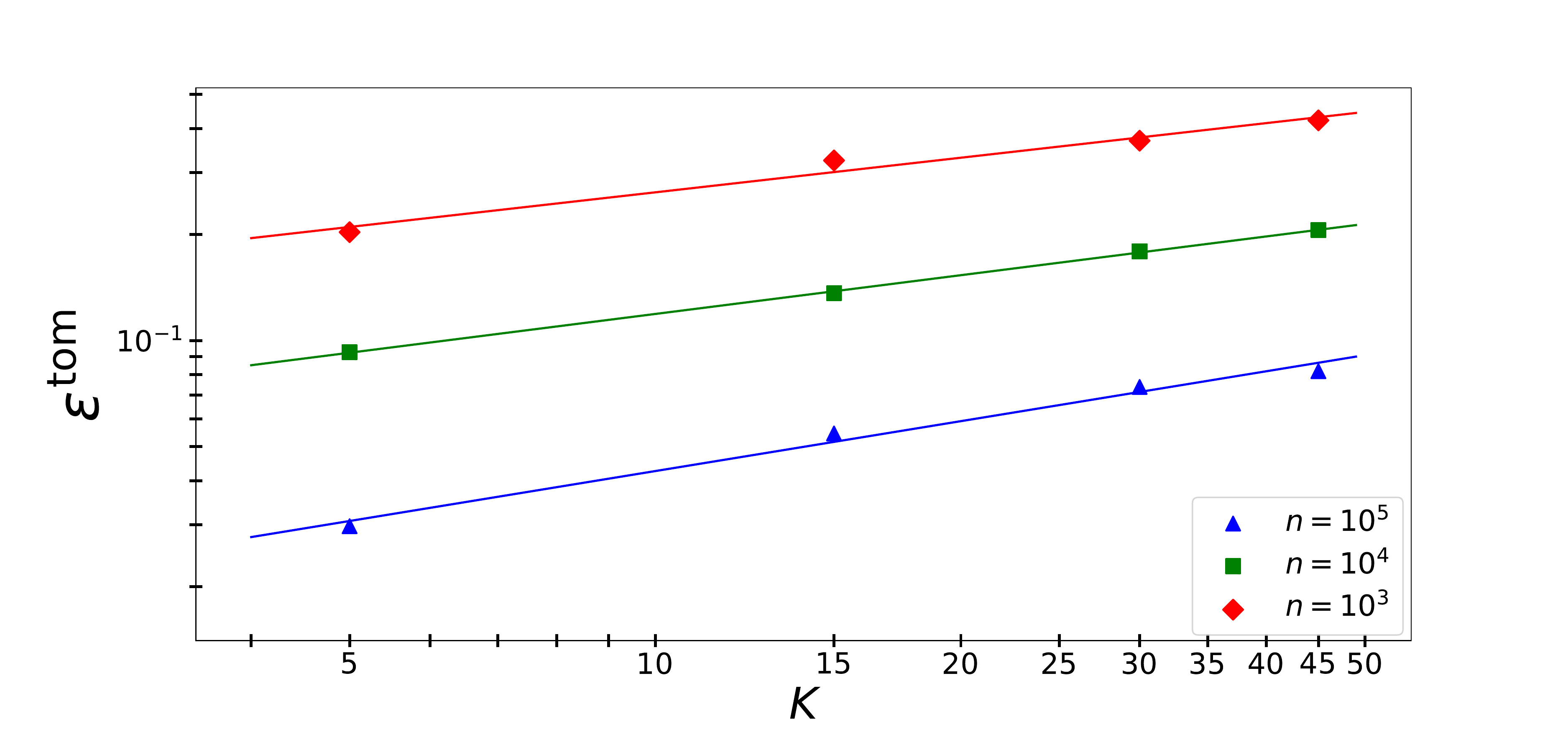}
\caption{Average
error~\eqref{tomographic-error-average-supplemental} for the
tomographic inference of the quantum dynamical map $\Phi_S(t)$ vs.
the total number of projective measurements used (top panel), vs.
the number $K$ of time moments analyzed, $t_i = i$ (bottom
panel).} \label{figure-9-supplemental}
\end{figure}

The standard quantum-process tomography exploits an ensemble of
identically prepared quantum systems corresponding to a given
experimental setting (see, e.g., the review~\cite{knee-2018}).
Suppose the total number of available projective measurements is
$n$. As we are interested in reconstructing $K$ channels
$\{\Phi_S(t_i)\}_{i=1}^K$ with the minimal possible average error,
the number of projective measurements per each channel equals
$\frac{n}{K}$. The theory of process
tomography~\cite{bogdanov-2013,haah-2017} predicts the
reconstruction error $\epsilon \sim \frac{d_S^4}{\sqrt{n/K}}$ in
this case. In what follows, we confirm this prediction
numerically.

Ref.~\cite{knee-2018} proposes an algorithm that maximizes the
likelihood for the observed measurement outcomes and provides a
legitimate (trace preserving and completely positive) estimate for
$\Phi_S^{\rm tom}(t_i)$. Dealing with a qubit dynamical map
($d_S=2$), one needs to prepare the system in one of four pure
initial states $\{ \varrho_S^{(j)}(0)\}_{j=1}^4$, where
$\varrho_S^{(1)}(0) = \ket{0}\bra{0}$, $\varrho_S^{(2)}(0) =
\ket{1}\bra{1}$, $\varrho_S^{(3)}(0) = \frac{1}{2} (\ket{0} +
\ket{1}) (\bra{0} + \bra{1})$, $\varrho_S^{(4)}(0) = \frac{1}{2}
(\ket{0} + i \ket{1}) (\bra{0} - i \bra{1})$. Then the randomly
chosen state $\varrho_S^{(j)}(0)$ is evolved through the channel
$\Phi_S^{\rm tom}(t_i)$ and is measured with the help of an
8-outcome positive operator-valued measure (POVM) with effects
$\{F_k\}_{k=1}^8$, where $F_{2m-1} = \frac{1}{4}
\varrho_S^{(m)}(0)$ and $F_{2m} = \frac{1}{4} [I -
\varrho_S^{(m)}(0)]$, $m=1,\ldots,4$. After the measurement
outcome is read out, the environment should be reset to the
initial (thermal equilibrium) state and the system should be again
prepared in one of the states $\{ \varrho_S^{(j)}(0)\}_{j=1}^4$.
This is a challenge in real experimental setup (especially in the
case of strong coupling between the system and environment) and a
disadvantage as compared to our proposed scheme of sequential
measurements with no environment resets. Suppose, however, that
the experiment is repeated $n/K$ times. This results in integers
$\{n_{jk}\}_{j=1,\ldots,4, \ k =1,\ldots,8}$, which quantify how
many times the outcome $k$ is observed provided the system is
prepared in the state $\varrho_S^{(j)}(0)$. Clearly, the relative
frequencies $\frac{4 K n_{jk}}{n}$ tend to probabilities $p_{jk} =
{\rm tr} \Big[ \Phi_S^{\rm tom}(t_i)[\varrho_S^{(j)}(0)] F_k
\Big]$ if $n\rightarrow \infty$. The authors of
Ref.~\cite{knee-2018} maximize the likelihood $\prod_{jk}
p_{jk}^{n_{jk}}$ with respect to $\Phi_S(t_i)$ and find the best
estimate $\Phi_S^{\rm tom}(t_i)$ for the quantum channel. We use
the solver in Ref.~\cite{knee-2018} to find $\Phi_S^{\rm
tom}(t_i)$ for a given number of measurements $n/K$.

Suppose the exact dynamical map $\Phi_S(t)$ is known, then the
average reconstruction error equals
\begin{equation} \label{tomographic-error-average-supplemental}
\varepsilon^{\rm tom}(\{\Phi_S(t_i)\}_{i=1}^K) = \frac{1}{2K}
\sum_{i=1}^{K} \| \Omega_{\Phi_S^{\rm tom}(t_i)} -
\Omega_{\Phi_S^{\rm exact}(t_i)} \|_1.
\end{equation}

\noindent Fig.~\ref{figure-9-supplemental} shows the error
$\varepsilon^{\rm tom}(\{\Phi_S(t_i)\}_{i=1}^K)$ scales as
$\sqrt{K / n}$, which is in strong contrast to
Fig.~\ref{figure-7-supplemental}.

\subsection{Details on the coherent control}

Within the process tensor formalism for Markovian embedding, we
are able to describe the action of coherent control pulses on the
system experiencing a non-Markovian dynamics. Suppose the system
is subjected to a quick unitary transformation $\varrho_S(t')
\rightarrow V\varrho_S(t') V^{\dag}$ at time moment $t'$. The
corresponding tensor diagram is depicted in
Fig.~\ref{figure-11-supplemental}.

The estimated system dynamics after the operation is given by the
equation $\varrho_S(t) = {\rm tr}_{ER} \big\{ \exp[(t-t'){\cal
L}_{S+ER}] \varrho_{S+ER}(t')\big\}$.
Fig.~\ref{figure-12-supplemental} illustrates the qubit evolution
for $V = \sigma_x$ and $t'=20$. The exact dynamics and the
estimated dynamics are in good agreement with each other.

Suppose the full process tomography is performed for time moments
$t_1, t_2, \ldots, t_K$ and the maps $\Phi_S(t_1), \Phi_S(t_2),
\ldots, \Phi_S(t_K)$ are reconstructed precisely. The dynamics
within the time interval $[t_i,t_{i+1}]$ is given by the
intermediate map $\Lambda(t_{i+1},t_i) =
\Phi_S(t_{i+1})\Phi_S^{-1}(t_i)$. Denoting $t_0=0$, we note that
$\Phi_S(t_j) = \Lambda(t_j,t_{j-1}) \cdots \Lambda(t_2,t_1)
\Lambda(t_1,t_0) =: \bigcirc_{i = 0}^{j-1} \Lambda(t_{i+1},t_i)$,
i.e., the dynamics is described by concatenation of intermediate
maps. If a coherent control gate $V$ is applied at time moment
$t_m$, then the concatenation approach yields
\begin{equation} \label{int-dyn}
\varrho_S(t_l) = \left\{ \begin{array}{ll}
\bigcirc_{i = 0}^{l-1} \Lambda(t_{i+1},t_i) [\varrho_S(0)]  = \Phi_S(t_l) [\varrho_S(0)] & \text{if~~} l < m,\\
\bigcirc_{i = m}^{l-1} \Lambda(t_{i+1},t_i) \Big[ V \left( \bigcirc_{j = 0}^{m-1} \Lambda(t_{j+1},t_j) [\varrho_S(0)] \right) V^{\dag} \Big] = \Phi_S(t_l) \Phi_S^{-1}(t_m) \Big[ V \Phi_S(t_m) [\varrho_S(0)] V^{\dag} \Big] & \text{if~~} l \geq m. \\
\end{array} \right.
\end{equation}

\noindent This approach results in the dynamics depicted in
Fig.~\ref{figure-12-supplemental} by dots. Clearly,
Eq.~\eqref{int-dyn} is not able to reproduce the system dynamics
after the control gate is applied because
$\Phi_S(t)\Phi_S(t')^{-1}[V\varrho_S(t') V^{\dag}] \neq
\varrho_S(t)$ for $t>t'$ due to the system-environment
correlations~\cite{gessner-2011,rivas-2014,milz-2019}.

Non-monotonicity of the trace distance $\frac{1}{2}\|\varrho_S'(t)
- \varrho_S''(t)\|_1$ for some initial states $\varrho_S'(0)$ and
$\varrho_S''(0)$ is a clear indication of
non-Markovianity~\cite{BLP}, and the learned Markovian embedding
reproduces such a non-monotonic behavior quite
well~\cite{lvgf-2019}.

\begin{figure}[H]
\centering
\includegraphics[width=12cm]{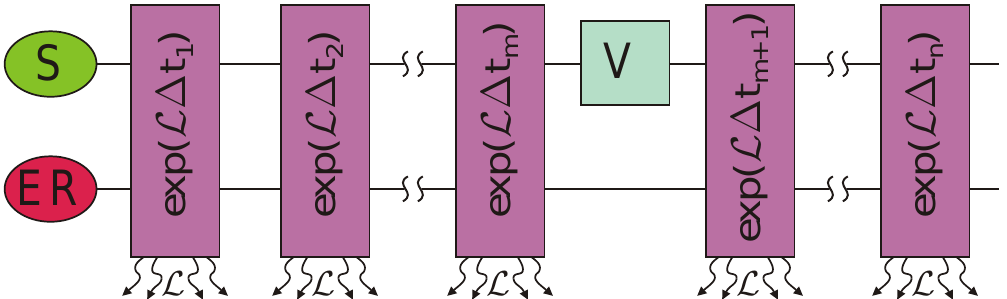}
\caption{Process tensor formalism is compatible with a coherent
control gate $V$ applied to the system at time $t'=t_m$.}
\label{figure-11-supplemental}
\end{figure}

\begin{figure}[H]
\centering
\includegraphics[width=12cm]{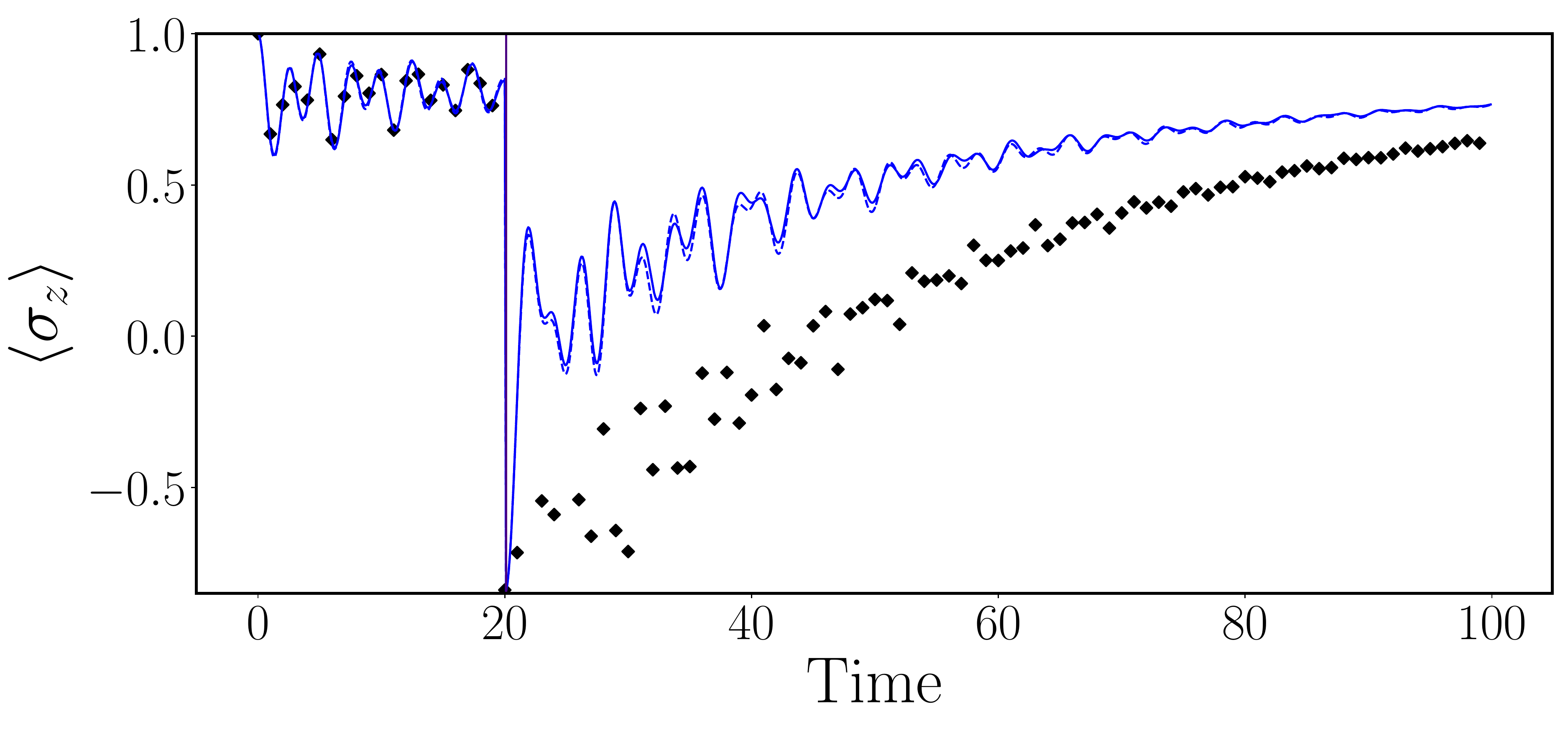}
\caption{Non-Markovian qubit dynamics with a quick control gate $V
= \sigma_x$ applied at $t'=20$: exact solution (solid line),
estimated solution within the Markovian embedding approach (dotted
line), solution~\eqref{int-dyn} within the full process tomography
approach (dots).} \label{figure-12-supplemental}
\end{figure}

\end{widetext}


\begin{thebibliography}{99}
\providecommand{\natexlab}[1]{#1}
\providecommand{\url}[1]{\texttt{#1}} \expandafter\ifx\csname
urlstyle\endcsname\relax
  \providecommand{\doi}[1]{doi: #1}\else
  \providecommand{\doi}{doi: \begingroup \urlstyle{rm}\Url}\fi

\bibitem{takahashi-2008}
S. Takahashi, R. Hanson, J. van Tol, M.~S. Sherwin, and D.~D.
Awschalom, Quenching spin decoherence in diamond through spin bath
polarization, Phys. Rev. Lett. {\bf 101}, 047601 (2008).

\bibitem{valkunas-2013}
L.~Valkunas, D.~Abramavicius, and T.~Mancal, \emph{Molecular
Excitation Dynamics and Relaxation: Quantum Theory and
Spectroscopy} (Wiley, New York, 2013).

\bibitem{degen-2017}
C.~L. Degen, F.~Reinhard, and P.~Cappellaro, Quantum sensing, Rev.
Mod. Phys. {\bf 89}, 035002 (2017).

\bibitem{wilde-2017}
M.~M. Wilde, \emph{Quantum Information Theory} (Cambridge
University Press, 2017).

\bibitem{nielsen-2000}
M.~A. Nielsen and I.~L. Chuang, \emph{Quantum Computation and
Quantum Information} (Cambridge University Press, Cambridge,
England, 2000).

\bibitem{schoeller-2018}
H. Schoeller, Dynamics of open quantum systems, arXiv:1802.10014
(2018).

\bibitem{piilo-2011}
B.-H. Liu, L. Li, Y.-F. Huang, C.-F. Li, G.-C. Guo, E.-M. Laine,
H.-P. Breuer, and J. Piilo, Experimental control of the transition
from Markovian to non-Markovian dynamics of open quantum systems,
Nat. Phys. {\bf 7}, 931 (2011).

\bibitem{cirac-2011}
C. Navarrete-Benlloch, I. de Vega, D. Porras, and J. I. Cirac,
Simulating quantum-optical phenomena with cold atoms in optical
lattices, New J. Phys. {\bf 13}, 023024 (2011).

\bibitem{ma-2012}
J. Ma, Z. Sun, X. Wang, and F. Nori, Entanglement dynamics of two
qubits in a common bath, Phys. Rev. A {\bf 85}, 062323 (2012).

\bibitem{hoope-2012}
U. Hoeppe, C. Wolff, J. K\"{u}chenmeister, J. Niegemann, M.
Drescher, H. Benner, and K. Busch, Direct observation of
non-Markovian radiation dynamics in 3D bulk photonic crystals,
Phys. Rev. Lett. {\bf 108}, 043603 (2012).

\bibitem{yang-2013}
W. L. Yang, J.-H. An, C. Zhang, M. Feng, and C. H. Oh,
Preservation of quantum correlation between separated
nitrogen-vacancy centers embedded in photonic-crystal cavities,
Phys. Rev. A {\bf 87}, 022312 (2013).

\bibitem{hughes-2015}
K. Roy-Choudhury and S. Hughes, Spontaneous emission from a
quantum dot in a structured photonic reservoir: phonon-mediated
breakdown of Fermi's golden rule, Optica {\bf 2}, 434 (2015).

\bibitem{eisert-2015}
S.~Gr{\"o}blacher, A.~Trubarov, N.~Prigge, G.~D.~Cole,
M.~Aspelmeyer, and J.~Eisert, Observation of non-Markovian
micromechanical Brownian motion, Nat. Commun. {\bf 6}, 7606
(2015).

\bibitem{cirac-2017}
A.~Gonz\'alez-Tudela and J.~I. Cirac, Quantum emitters in
two-dimensional structured reservoirs in the nonperturbative
regime, Phys. Rev. Lett. {\bf 119}, 143602 (2017).

\bibitem{wittemer-2018}
M. Wittemer, G. Clos, H.-P. Breuer, U. Warring, and T. Schaetz,
Measurement of quantum memory effects and its fundamental
limitations, Phys. Rev. A {\bf 97}, 020102(R) (2018).

\bibitem{wang-2018}
F. Wang, P.-Y. Hou, Y.-Y. Huang, W.-G. Zhang, X.-L. Ouyang, X.
Wang, X.-Z. Huang, H.-L. Zhang, L. He, X.-Y. Chang, and L.-M.
Duan, Observation of entanglement sudden death and rebirth by
controlling a solid-state spin bath, Phys. Rev. B {\bf 98}, 064306
(2018).

\bibitem{peng-2018}
S. Peng, X. Xu, K. Xu, P. Huang, P. Wang, X. Kong, X. Rong, F.
Shi, C. Duan, and J. Du, Observation of non-Markovianity at room
temperature by prolonging entanglement in solids, Science Bulletin
{\bf 63}, 336 (2018).

\bibitem{haase-2018}
J. F. Haase, P. J. Vetter, T. Unden, A. Smirne, J. Rosskopf, B.
Naydenov, A. Stacey, F. Jelezko, M. B. Plenio, and S. F. Huelga,
Controllable non-Markovianity for a spin qubit in diamond, Phys.
Rev. Lett. {\bf 121}, 060401 (2018).

\bibitem{mascherpa-2019}
F. Mascherpa, A. Smirne, A. D. Somoza, P. Fern\'{a}ndez-Acebal, S.
Donadi, D. Tamascelli, S. F. Huelga, and M. B. Plenio, Optimized
auxiliary oscillators for the simulation of general open quantum
systems, arXiv:1904.04822 [quant-ph].

\bibitem{strathearn-2018}
A. Strathearn, P. Kirton, D. Kilda, J. Keeling, and B.~W. Lovett,
Efficient non-Markovian quantum dynamics using time-evolving
matrix product operators, Nat. Commun. {\bf 9}, 3322 (2018).

\bibitem{pollock-2019}
M.~R. {J{\o}rgensen} and F.~A. {Pollock}, Exploiting the causal
tensor network structure of quantum processes to efficiently
simulate non-Markovian path integrals, Phys. Rev. Lett. {\bf 123},
240602 (2019).

\bibitem{altaisky-2017}
M. V. Altaisky, N. N. Zolnikova, N. E. Kaputkina, V. A. Krylov,
Yu. E. Lozovik, and N. S. Dattani, Entanglement in a quantum
neural network based on quantum dots, Photonics and Nanostructures
-- Fundamentals and Applications {\bf 24}, 24 (2017).

\bibitem{paris-2018}
M. Bina, F. Grasselli, and M.~G.~A. Paris, Continuous-variable
quantum probes for structured environments, Phys. Rev. A {\bf 97},
012125 (2018).

\bibitem{bennink-2019}
R. S. Bennink and P. Lougovski, Quantum process identification: a
method for characterizing non-markovian quantum dynamics, New J.
Phys. {\bf 21}, 083013 (2019).

\bibitem{gks-1976}
V. Gorini, A. Kossakowski, and E.~C.~G. Sudarshan, Completely
positive dynamical semigroups of n-level systems, J. Math. Phys.
(N.Y.) {\bf 17}, 821 (1976).

\bibitem{lindblad-1976}
G.~Lindblad, On the generators of quantum dynamical semigroups,
Commun. Math. Phys. {\bf 48}, 119 (1976).

\bibitem{Tomography}
I.~L. Chuang and M.~A. Nielsen, Prescription for experimental
determination of the dynamics of a quantum black box, J. Mod. Opt.
{\bf 44}, 2455 (1997).

\bibitem{howard-2006}
M.~Howard, J.~Twamley, C.~Wittmann, T.~Gaebel, F.~Jelezko, and
J.~Wrachtrup, Quantum process tomography and Lindblad estimation
of a solid-state qubit, New J. Phys. {\bf 8}, 33 (2006).

\bibitem{de-vega-2017}
I. de~Vega and D. Alonso, Dynamics of non-Markovian open quantum
systems, Rev. Mod. Phys. {\bf 89}, 015001 (2017).

\bibitem{li-2018}
L.~Li, M.~J.~W. Hall, and H.~M. Wiseman, Concepts of quantum
non-Markovianity: A hierarchy, Phys. Rep. {\bf 759}, 1 (2018).

\bibitem{fc-2018}
S.~N. Filippov and D. Chru\'{s}ci\'{n}ski, Time deformations of
master equations, Phys. Rev. A {\bf 98}, 022123 (2018).

\bibitem{bogdanov-2013}
Yu.~I. Bogdanov, A.~A. Kalinkin, S.~P. Kulik, E.~V. Moreva, and
V.~A. Shershulin, Quantum polarization transformations in
anisotropic dispersive media, New J. Phys. {\bf 15}, 035012
(2013).

\bibitem{haah-2017}
J.~Haah, A.~W. Harrow, Z.~Ji, X.~Wu, and N.~Yu, Sample-optimal
tomography of quantum states, IEEE Trans. Inf. Theory {\bf 63},
5628 (2017).

\bibitem{Cerrilo-2014}
J. Cerrillo and J. Cao, Non-Markovian dynamical maps: Numerical
processing of open quantum trajectories, Phys. Rev. Lett. {\bf
112}, 110401 (2014).

\bibitem{TTM}
A. Gelzinis, E. Rybakovas, and L. Valkunas, Applicability of
transfer tensor method for open quantum system dynamics, J. Chem.
Phys. {\bf 147}, 234108 (2017).

\bibitem{TTMTomography}
F.~A. Pollock and K. Modi, Tomographically reconstructed master
equations for any open quantum dynamics, Quantum {\bf 2}, 76
(2018).

\bibitem{nakajima-1958}
S. Nakajima, On quantum theory of transport phenomena: Steady
diffusion, Prog. Theor. Phys. {\bf 20}, 948 (1958).

\bibitem{zwanzig-1960}
R. Zwanzig, Ensemble method in the theory of irreversibility, J.
Chem. Phys. {\bf 33}, 1338 (1960).

\bibitem{RNN}
L. Banchi, E. Grant, A. Rocchetto, and S. Severini, Modelling
non-Markovian quantum processes with recurrent neural networks,
New J. Phys. {\bf 20}, 123030 (2018).

\bibitem{MarkovEmb}
A.~A. Budini, Embedding non-Markovian quantum collisional models
into bipartite Markovian dynamics, Phys. Rev. A {\bf 88}, 032115
(2013).

\bibitem{xue-2015}
S. Xue, M.~R. James, A. Shabani, V. Ugrinovskii, and I.~R.
Petersen, Quantum filter for a class of non-Markovian quantum
systems, in {\it 54th IEEE Conference on Decision and Control
(Osaka, Japan)} (IEEE, New York, 2015), pp. 7096--7100.

\bibitem{xue-2017}
S. Xue, T. Nguyen, M.~R. James, A. Shabani, V. Ugrinovskii, and
I.~R. Petersen, Modelling and filtering for non-Markovian quantum
systems, arXiv:1704.00986 (2017).

\bibitem{campbell-2018}
S. Campbell, F. Ciccarello, G.~M. Palma, and B. Vacchini,
System-environment correlations and Markovian embedding of quantum
non-Markovian dynamics, Phys. Rev. A {\bf 98}, 012142 (2018).

\bibitem{luchnikov-2019}
I.~A. Luchnikov, S.~V. Vintskevich, H.~Ouerdane, and S.~N.
Filippov, Simulation complexity of open quantum dynamics:
Connection with tensor networks, Phys. Rev. Lett. {\bf 122},
160401 (2019).

\bibitem{imamoglu-1994}
A.~Imamoglu, Stochastic wave-function approach to non-Markovian
systems, Phys. Rev. A {\bf 50}, 3650 (1994).

\bibitem{garraway-1997}
B.~M. Garraway, Nonperturbative decay of an atomic system in a
cavity, Phys. Rev. A {\bf 55}, 2290 (1997).

\bibitem{mazzola-2009}
L.~Mazzola, S.~Maniscalco, J.~Piilo, K.-A. Suominen, and B.~M.
Garraway, Pseudomodes as an effective description of memory:
Non-Markovian dynamics of two-state systems in structured
reservoirs, Phys. Rev. A {\bf 80}, 012104 (2009).

\bibitem{iles-smith-2014}
J. Iles-Smith, N. Lambert, and A. Nazir, Environmental dynamics,
correlations, and the emergence of noncanonical equilibrium states
in open quantum systems, Phys. Rev. A {\bf 90}, 032114 (2014).

\bibitem{iles-smith-2016}
J. Iles-Smith, A. G. Dijkstra, N. Lambert, and A. Nazir, Energy
transfer in structured and unstructured environments: Master
equations beyond the Born-Markov approximations, J. Chem. Phys.
{\bf 144}, 044110 (2016).

\bibitem{tamascelli-2018}
D. Tamascelli, A. Smirne, S. F. Huelga, and M. B. Plenio,
Nonperturbative treatment of non-Markovian dynamics of open
quantum systems, Phys. Rev. Lett. {\bf 120}, 030402 (2018).

\bibitem{polyakov-2019}
E. A. Polyakov and A. N. Rubtsov, Dressed quantum trajectories:
novel approach to the non-Markovian dynamics of open quantum
systems on a wide time scale, New J. Phys. {\bf 21}, 063004
(2019).

\bibitem{carleo-2019}
G. Carleo, I. Cirac, K. Cranmer, L. Daudet, M. Schuld, N. Tishby,
L. Vogt-Maranto, and L. Zdeborov\'{a}, Machine learning and the
physical sciences, Rev. Mod. Phys. {\bf 91}, 045002 (2019).

\bibitem{supplemental}
See Supplemental Material for details on the derivation of the
likelihood function gradient, details on the learning algorithm
(including overfitting and the hyperparameter $d_{\rm ER}$),
details on the data generation, details on the error estimation
(including the variational Bayesian inference approach and a
comparison with the full process tomography), and details on the
coherent control, which includes
Refs.~\cite{fsp-2020,molchanov-2017,kingma-2013,filippov-jms-2019,knee-2018,BLP,lvgf-2019}.

\bibitem{fsp-2020}
S. N. Filippov, G. N. Semin, and A. N. Pechen, Quantum master
equations for a system interacting with a quantum gas in the
low-density limit and for the semiclassical collision model, Phys.
Rev. A {\bf 101}, 012114 (2020).

\bibitem{molchanov-2017}
D. Molchanov, A. Ashukha, and D. Vetrov, Variational dropout
sparsifies deep neural networks, in \textit{Proceedings of the
34th International Conference on Machine Learning}, edited by D.
Precup and Y. W. Teh (PMLR, Cambridge, MA, 2017), Vol. 70, pp.
2498--2507,
\url{http://proceedings.mlr.press/v70/molchanov17a.html}.

\bibitem{kingma-2013} D. P. Kingma and M. Welling, Auto-encoding variational
Bayes, in \textit{Proceedings of the 2nd International Conference
on Learning Representations} (ICLR 2014), arXiv:1312.6114
[stat.ML].

\bibitem{filippov-jms-2019}
S. N. Filippov, Quantum mappings and characterization of entangled
quantum states, J. Math. Sci. {\bf 241}, 210 (2019).

\bibitem{knee-2018}
G. C. Knee, E. Bolduc, J. Leach, and E. M. Gauger, Quantum process
tomography via completely positive and trace-preserving
projection, Phys. Rev. A {\bf 98}, 062336 (2018).

\bibitem{BLP}
E.-M. Laine, J. Piilo, and H.-P. Breuer, Measure for the
non-Markovianity of quantum processes, Phys. Rev. A {\bf 81},
062115 (2010).

\bibitem{lvgf-2019}
I. A. Luchnikov, S. V. Vintskevich, D. A. Grigoriev, and S. N.
Filippov, Machine learning non-Markovian quantum dynamics,
arXiv:1902.07019v2 [quant-ph].

\bibitem{shrapnel-2018}
S. Shrapnel, F. Costa, and G. Milburn, Quantum Markovianity as a
supervised learning task, Int. J. Quantum Inf. {\bf 16}, 1840010
(2018).

\bibitem{lindblad-1979}
G. Lindblad, Non-Markovian quantum stochastic processes and their
entropy, Commun. Math. Phys. {\bf 65}, 281 (1979).

\bibitem{pollock-2018}
F.~A. Pollock, C. Rodr\'{\i}guez-Rosario, T. Frauenheim, M.
Paternostro, and K. Modi, Non-Markovian quantum processes:
Complete framework and efficient characterization, Phys. Rev. A
{\bf 97}, 012127 (2018).

\bibitem{modi-2018}
F.~A. Pollock, C. Rodr\'{\i}guez-Rosario, T. Frauenheim, M.
Paternostro, and K. Modi, Operational Markov condition for quantum
processes, Phys. Rev. Lett. {\bf 120}, 040405 (2018).

\bibitem{budini-2018}
A.~A. Budini, Quantum non-Markovian processes break conditional
past-future independence, Phys. Rev. Lett. {\bf 121}, 240401
(2018).

\bibitem{taranto-2019}
P. Taranto, F.~A. Pollock, S. Milz, M. Tomamichel, and K. Modi,
Quantum Markov order, Phys. Rev. Lett. {\bf 122}, 140401 (2019).

\bibitem{costa-2016}
F. Costa and S. Shrapnel, Quantum causal modelling, New J. Phys.
{\bf 18}, 063032 (2016).

\bibitem{milz-2017}
S. Milz, F.~A. Pollock, and K. Modi, An introduction to
operational quantum dynamics, Open Syst. Inf. Dyn. {\bf 24},
1740016 (2017).

\bibitem{chiribella-2009}
G. Chiribella, G.~M. D'Ariano, and P. Perinotti, Theoretical
framework for quantum networks, Phys. Rev. A {\bf 80}, 022339
(2009).

\bibitem{milz-2018}
S. Milz, F.~A. Pollock, and K. Modi, Reconstructing non-Markovian
quantum dynamics with limited control, Phys. Rev. A {\bf 98},
012108 (2018).

\bibitem{luchnikov-2017}
I.~A. Luchnikov and S.~N. Filippov, Quantum evolution in the
stroboscopic limit of repeated measurements, Phys. Rev. A {\bf
95}, 022113 (2017).

\bibitem{comment}
$\Phi^{\dag}$ is dual to $\Phi$ if ${\rm tr} \big[X \Phi[Y] \big]
= {\rm tr} \big[ \Phi^{\dag}[X] Y \big]$ for all $X$ and $Y$.

\bibitem{molmer-2013}
S. Gammelmark, B. Julsgaard, and K. M{\o}lmer, Past quantum states
of a monitored system, Phys. Rev. Lett. {\bf 111}, 160401 (2013).

\bibitem{Bishop}
C. M. Bishop, \emph{Pattern Recognition and Machine Learning}
(Springer, New York, 2006).

\bibitem{CVX}
S.~Boyd, L.~Vandenberghe, \emph{Convex Optimization} (Cambridge
University Press, Cambridge, England, 2004).

\bibitem{Holevo}
A.~S. Holevo, \emph{Quantum Systems, Channels, Information: A
Mathematical Introduction} (De Gruyter, Berlin, 2012).

\bibitem{rau-1963}
J.~Rau, Relaxation phenomena in spin and harmonic oscillator
systems, Phys. Rev. {\bf 129}, 1880 (1963).

\bibitem{scarani-2002}
V. Scarani, M. Ziman, P. \v{S}telmachovi\v{c}, N. Gisin, and V.
Bu\v{z}ek, Thermalizing quantum machines: Dissipation and
entanglement, Phys. Rev. Lett. {\bf 88}, 097905 (2002).

\bibitem{filippov-2017}
S.~N. Filippov, J.~Piilo, S.~Maniscalco, and M.~Ziman,
Divisibility of quantum dynamical maps and collision models, Phys.
Rev. A {\bf 96}, 032111 (2017).

\bibitem{jain-2017}
P. Jain and P. Kar, Non-convex optimization for machine learning,
Found. Trends Mach. Learn. {\bf 10}, 142 (2017).

\bibitem{grigoriev_github}
I. A. Luchnikov, S. V. Vintskevich, D. A. Grigoriev, and S. N.
Filippov, Machine learning of Markovian embedding for
non-Markovian quantum dynamics, GitHub repository (2019),
\href{https://github.com/GrigorievDmitry/Machine-learning-of-Markovian-embedding-for-non-Markovian-quantum-dynamics}{\tt{https://github.com/GrigorievDmitry/Machine
  learning of Markovian embedding for non-Markovian quantum dynamics}}.

\bibitem{Adam}
D.~P. Kingma and J. Ba, Adam: A method for stochastic
optimization, arXiv:1412.6980 [cs.LG] (2014).

\bibitem{lorenzo-2017}
S. Lorenzo, F. Ciccarello, and G.~M.~Palma, Composite quantum
collision models, Phys. Rev. A {\bf 96}, 032107 (2017).

\bibitem{gessner-2011}
M. Gessner and H.-P. Breuer, Detecting nonclassical
system-environment correlations by local operations, Phys. Rev.
Lett. {\bf 107}, 180402 (2011).

\bibitem{rivas-2014}
\'{A}. Rivas, S. F. Huelga, and M. B. Plenio, Quantum
non-Markovianity: Characterization, quantification and detection,
Rep. Prog. Phys. {\bf 77}, 094001 (2014).

\bibitem{milz-2019}
S. Milz, M.~S. Kim, F.~A. Pollock, and K.~Modi, Completely
positive divisibility does not mean Markovianity, Phys. Rev. Lett.
{\bf 123}, 040401 (2019).

\end{thebibliography}
\end{document}